\documentclass[12pt,thmsa]{article}
\usepackage{amssymb}
\usepackage{epsfig}
\usepackage{sw20jart}

\input{tcilatex}
\begin{document}

\title{Self-Similar Evaporation of a Rigidly-Rotating Cosmic String Loop}
\author{Malcolm Anderson \\
Department of Mathematics\\
Universiti Brunei Darussalam\\
Jalan Tungku Link, Gadong BE 1410\\
Negara Brunei Darussalam}
\date{}
\maketitle

\begin{abstract}
\noindent The gravitational back-reaction on a certain type of
rigidly-rotating cosmic string loop, first discovered by Allen, Casper and
Ottewill, is studied at the level of the weak-field approximation. The
near-field metric perturbations are calculated and used to construct the
self-acceleration vector of the loop. Although the acceleration vector is
divergent at the two kink points on the loop, its net effect on the
trajectory over a single oscillation period turns out to be finite. The net
back-reaction on the loop over a single period is calculated using a method
due to Quashnock and Spergel, and is shown to induce a uniform shrinkage of
the loop while preserving its original shape. The loop therefore evolves by
self-similar evaporation.\bigskip

Short Title: Evaporation of a Cosmic String Loop\bigskip

\noindent PACS\ numbers: 04.25.Nx, 98.80.Cq
\end{abstract}

\section{Introduction}

Cosmic strings are thin filaments of Higgs field energy that may have played
a role in the formation of large-scale structure in the early Universe. The
dynamics of cosmic strings in Minkowski and Robertson-Walker backgrounds has
been studied extensively \cite{Kibb1, Kibb2, Thomp, Vilenkin, Vega1, Vega2},
and their possible cosmogonic effects are well known (see \cite{Vile-Shell}
for a review), but their gravitational properties remain relatively
unexplored. In particular, the motion of a string loop under the action of
its own gravitational field is described by a non-linear system of equations
that can normally be solved only by numerical methods, even in the case of a
zero-thickness string in the weak-field limit \cite{Quash}.

Quite apart from the potential cosmological ramifications, a better
understanding of the gravitational field and back-reaction effects of
general cosmic string configurations is important for the theoretical
development of general relativity. Zero-thickness cosmic strings form a
class of line singularities that are natural higher-dimensional analogues of
black holes, and it remains an open question whether they can be
satisfactorily incorporated into the framework of general relativity. Areas
that are usually regarded as problematic include the definition of an
invariant characterisation of zero-thickness cosmic strings, the development
of a suitable distributional formalism for line singularities, and the
demonstration that zero-thickness cosmic strings are the unique limits of
well-behaved finite-thickness solutions \cite{Geroch}. These issues can all
be resolved in the few known cases of exact self-gravitating solutions \cite
{Vick1, Vick2, Garf1, Anderson}, but a general resolution awaits further
work.

The uncertain status of line singularities in general relativity is
underscored by the back-reaction problem. Almost nothing is known about the
motion of a zero-thickness cosmic string under the action of its
self-gravity. Although it can be shown that the self-gravity of a GUT string
would almost everywhere be small enough to justify a weak-field treatment
\cite{Quash}, even the weak-field dynamics is complicated and therefore
little studied. Furthermore, flat-space string solutions commonly support
pathological features -- kinks and cusps -- where the weak-field
approximation is known to break down \cite{Garf2, Vachaspati}. Much work
needs to be done to justify the claim that zero-thickness cosmic strings can
move freely and self-consistently under their own gravity, and are therefore
compatible with general relativity.

In this paper I calculate the gravitational field and back-reaction force on
a certain type of rigidly-rotating cosmic string loop -- the
Allen-Casper-Ottewill solution \cite{ACO1} -- in the weak-field limit, and
show that the loop evolves by self-similar shrinkage. This is the first
explicit back-reaction calculation to be published for a cosmic string loop,
as all previous calculations have relied on numerical approximation \cite
{Quash}. The importance of the results described below lies not so much in
the simple behaviour of the loop, but in the detailed features of the
gravitational effects involved. In particular, the loop supports a pair of
kinks near which the back-reaction force diverges but nonetheless remains
integrable.

\section{Preliminaries}

A zero-thickness cosmic string is a line singularity which as it moves
traces out a two-dimensional surface $\mathbf{T}$ whose dynamics and
stress-energy content are governed by the Nambu-Goto action. In what
follows, $x^{a}\equiv [x^{0},x^{1},x^{2},x^{3}]$ are local coordinates on
the four-dimensional background spacetime $(\mathbf{M},g_{ab})$, and the
metric tensor $g_{ab}$ has signature $(+,-,-,-)$, so that timelike vectors
have positive norm. The world sheet $\mathbf{T}$ of a general zero-thickness
cosmic string is then described parametrically by a set of equations of the
form $x^{a}=X^{a}(\zeta ^{A})$, where the parameters $\zeta ^{A}\equiv
(\zeta ^{0},\zeta ^{1})$ are commonly referred to as world-sheet or gauge
coordinates.

The intrinsic two-metric induced on $\mathbf{T}$ by a given choice of gauge
coordinates is:
\begin{equation}
\gamma _{AB}=g_{ab}\,X^{a},_{A}\,X^{b},_{B}
\end{equation}
where $X^{a},_{A}$ is shorthand for $\partial X^{a}/\partial \zeta ^{A}$.
For a string composed of non-tachyonic material $\gamma _{AB}$ is
non-degenerate with signature $(+,-)$ almost everywhere on $\mathbf{T}$,
although $\gamma _{AB}$ may be degenerate at isolated points called cusps.
If $\gamma $ denotes $|\det (\gamma _{AB})|$ the Nambu-Goto action \cite
{Nambu, Goto} has the form
\begin{equation}
I=-\mu \int \gamma ^{1/2}d^{2}\zeta  \label{action}
\end{equation}
where $\mu $ is the mass per unit length of the string.

The Lagrangian density $\mathcal{L}\equiv -\mu \gamma ^{1/2}$ in (\ref
{action}) is a functional of the position functions $X^{a}$ and their gauge
derivatives, and of the background metric $g_{ab}$. The equation of motion
of the corresponding string is recovered by setting the functional
derivative $\delta \mathcal{L}/\delta X^{a}$ equal to zero, while the
stress-energy tensor $T^{ab}$ at a general point on the world sheet $\mathbf{%
T}$ is $-2g^{-1/2}\delta \mathcal{L}/\delta g_{ab}$, where $g\equiv |\det
(g_{ab})|$. For the purposes of calculation, it is convenient to first fix
the gauge so that $\gamma _{AB}$ is everywhere proportional to the Minkowski
tensor $\eta _{AB}=$diag$\,(1,-1)$, a choice often referred to as the
standard gauge. The corresponding gauge coordinates are conventionally
written as $\zeta ^{0}\equiv \tau $ and $\zeta ^{1}\equiv \sigma $.

In the standard gauge, with $X_{\sigma }^{a}=\partial X^{a}/\partial \sigma $
and $X_{\tau }^{a}=\partial X^{a}/\partial \tau $, the string equation of
motion reads:
\begin{equation}
X^{a},_{\tau \tau }+\Gamma _{bc}^{a}\,X_{\tau }^{b}\,X_{\tau
}^{c}=X^{a},_{\sigma \sigma }+\Gamma _{bc}^{a}\,X_{\sigma }^{b}\,X_{\sigma
}^{c}  \label{eqmo}
\end{equation}
where $\Gamma _{bc}^{a}=\frac{1}{2}%
g^{ad}(g_{db},_{c}+g_{dc},_{b}-g_{bc},_{d})$ is the Christoffel symbol on
the background spacetime, evaluated at $x^{a}=X^{a}(\tau ,\sigma )$. Also,
the stress-energy tensor at a general spacetime point $x^{c}$ in the absence
of other material sources is:
\begin{equation}
T^{ab}(x^{c})=\mu g^{-1/2}\int (X_{\tau }^{a}\,X_{\tau }^{b}-X_{\sigma
}^{a}\,X_{\sigma }^{b})\,\delta ^{(4)}(x^{c}-X^{c})\,d\tau \,d\sigma
\label{SE}
\end{equation}
The strong-field version of the back-reaction problem is completely
specified by combining (\ref{eqmo}) and (\ref{SE}) with the Einstein
equation $G^{ab}=-8\pi T^{ab}$, where $G^{ab}=R^{ab}-\frac{1}{2}g^{ab}R$ is
the Einstein tensor.\footnote{%
The convention adopted here for the Riemann tensor is that
\[
R_{abcd}=\tfrac{1}{2}
(g_{ac},_{bd}+g_{bd},_{ac}-g_{ad},_{bc}-g_{bc},_{ad})+g_{ef}\,(\Gamma
_{ac}^{e}\Gamma _{bd}^{f}-\Gamma _{ad}^{e}\Gamma _{bc}^{f}).
\]
Also, geometrised units have been chosen, so that the gravitational constant
$G$ and the speed of light $c$ are set to $1$.}

To date, the only known solutions to the strong-field back-reaction problem
describe either infinite straight strings (with or without an envelope of
cylindrical gravitational waves \cite{Gott, Hiscock, Xanth, Garri}) or
infinite strings interacting with plane-fronted gravitational waves (the
so-called travelling-wave solutions \cite{Garf3, Frolov}). One of the
reasons why an explicit solution is possible in these cases is that the
world sheets are intrinsically flat and non-dissipative. A self-gravitating
string loop, by contrast, would be expected to continually radiate energy
and angular momentum, and there seems little prospect of generating
strong-field solutions of this type. The weak-field approximation is
therefore the only tool currently available for analysing loop evolution.

In the weak-field approximation $g_{ab}=\eta _{ab}+h_{ab}$, where $\eta
_{ab}=\,$diag$\,(1,-1,-1,-1)$ and the components of $h_{ab}$ are all small
in absolute value compared to $1$. To linear order in $h_{ab}$ the string
equation of motion (\ref{eqmo}) then reads
\begin{equation}
X^{a},_{\tau \tau }-X^{a},_{\sigma \sigma }=\alpha ^{a}  \label{eqmo2}
\end{equation}
where
\begin{equation}
\alpha ^{a}=(h_{b}^{a},_{c}-\tfrac{1}{2}\eta ^{ad}h_{bc},_{d})(X_{\sigma
}^{b}\,X_{\sigma }^{c}-X_{\tau }^{b}X_{\tau }^{c})  \label{accel}
\end{equation}
is the local four-acceleration of the string. Here, indices are everywhere
raised and lowered using the Minkowski tensor $\eta _{ab}$ and its inverse $%
\eta ^{ab}$. For future reference, $h\equiv h_{a}^{a}$ and inner products
such as $X_{\tau }^{2}$, $X_{\sigma }^{2}$ and $X_{\sigma }\,\cdot X_{\tau }$
are formed using $\eta _{ab}$.

If the harmonic gauge conditions $h_{a}^{b},_{b}=\frac{1}{2}h,_{a}$ are
imposed, the Einstein equation $G^{ab}=-8\pi T^{ab}$ reduces to $\square
h_{ab}=-16\pi S_{ab}$, where $S_{ab}=T_{ab}-\frac{1}{2}\eta _{ab}T_{c}^{c}$
and $\square \equiv \partial _{t}^{2}-\nabla ^{2}$ is the flat-space
d'Alembertian. With $x^{a}\equiv [t,\mathbf{x}]^{a}$ the standard solution
for $h_{ab}$ is:
\begin{equation}
h_{ab}(t,\mathbf{x})=-4\int \frac{S_{ab}(t^{\prime },\mathbf{x}^{\prime })}{|%
\mathbf{x}-\mathbf{x}^{\prime }|}\,d^{3}x^{\prime }  \label{weakh}
\end{equation}
where $t^{\prime }=t-|\mathbf{x}-\mathbf{x}^{\prime }|$ is the retarded time
at the source point $\mathbf{x}^{\prime }$.

When the source function $S_{ab}$ corresponding to the string stress-energy
tensor (\ref{SE}) is inserted into (\ref{weakh}) an integral over $\tau $, $%
\sigma $ and the three components of $\mathbf{x}^{\prime }$ results. The
distributional factor $\delta ^{(4)}(x^{\prime c}-X^{c})$ can be integrated
out by first transforming from $y^{a}\equiv [\tau ,\mathbf{x}^{\prime }]^{a}$
to $z^{a}\equiv [t^{\prime },\mathbf{x}^{\prime }]^{a}-X^{a}(\tau ,\sigma )$%
, with $\sigma $, $t$ and $\mathbf{x}$ fixed. The Jacobian of this
transformation is:
\begin{equation}
\left| \partial \,y^{a}/\partial \,z^{b}\right| =\left| \partial
\,z^{b}/\partial \,y^{a}\right| ^{-1}=|X_{\tau }^{0}-\mathbf{n}^{\prime
}\cdot \mathbf{X}_{\tau }|^{-1},
\end{equation}
where $\mathbf{n}^{\prime }=(\mathbf{x}-\mathbf{x}^{\prime })/|\mathbf{x}-%
\mathbf{x}^{\prime }|$ is the unit vector from the source point $\mathbf{x}%
^{\prime }$ to the field point $\mathbf{x}$, and $X^{a}\equiv [X^{0},\mathbf{%
X}]^{a}$. On integrating over $z^{a}$, $t^{\prime }$ is everywhere replaced
by $X^{0}$ and $\mathbf{x}^{\prime }$ by $\mathbf{X}$. In particular, the
gauge coordinate $\tau $ becomes an implicit function of $\sigma $ through
the light-cone condition
\begin{equation}
X^{0}(\tau ,\sigma )=t-|\mathbf{x}-\mathbf{X}(\tau ,\sigma )|.
\label{lightcon}
\end{equation}
The corresponding curve in the $\tau -\sigma $ plane, which represents the
intersection of the past light-cone of the field point with the string world
sheet, will be denoted by $\Gamma $.

Since $g^{1/2}=1$ to leading order in $h_{ab}$, the solution for $h_{ab}$
reads
\begin{equation}
h_{ab}(t,\mathbf{x})=-4\mu \int_{\Gamma }\frac{\Psi _{ab}}{|\mathbf{x}-%
\mathbf{X}|}\,|X_{\tau }^{0}-\mathbf{n}\cdot \mathbf{X}_{\tau }|^{-1}d\sigma
\label{weakh2}
\end{equation}
where now $\mathbf{n}=(\mathbf{x}-\mathbf{X})/|\mathbf{x}-\mathbf{X}|$ and
\begin{equation}
\Psi ^{ab}(\tau ,\sigma )=X_{\tau }^{a}\,X_{\tau }^{b}-X_{\sigma
}^{a}\,X_{\sigma }^{b}-\tfrac{1}{2}\,\eta ^{ab}\,(X_{\tau }^{2}-X_{\sigma
}^{2}).  \label{source}
\end{equation}
The equation of motion (\ref{eqmo2}) and the expression (\ref{weakh2}) for
the metric perturbations $h_{ab}$ together constitute the weak-field
back-reaction problem.

It should be stressed that the weak-field equation of motion (\ref{eqmo2})
holds only in a flat background, and in a particular gauge. Since questions
have been raised about the validity of this equation \cite{Cart2} it is
worthwhile taking the time to briefly review its derivation. The basic
assumption underlying the weak-field back-reaction formalism is that any
solution pair $(g_{ab},X^{a})$ to the strong-field problem represented by
equations (\ref{eqmo}) and (\ref{SE}) can -- for small values of the mass
per unit length $\mu $ at least -- be expanded as perturbative series of the
form
\begin{equation}
g_{ab}=\sum_{k=0}^{\infty }\mu ^{k}g_{ab}^{(k)}\text{\qquad and\qquad }%
X^{a}=\sum_{k=0}^{\infty }\mu ^{k}X_{(k)}^{a}  \label{expand}
\end{equation}
where the functions $g_{ab}^{(k)}$ and $X_{(k)}^{a}$ are independent of $\mu
$.

In the problem at hand, $g_{ab}^{(0)}=\eta _{ab}$ and $\mu
g_{ab}^{(1)}=h_{ab}$, while the position vector $X^{a}$ appearing in (\ref
{eqmo2}) should be read as $X_{(0)}^{a}+\mu X_{(1)}^{a}$. However, for
consistency the occurrences of $X^{a}$ in (\ref{accel}), (\ref{weakh2}) and (%
\ref{source}) are all truncated at $X_{(0)}^{a}$ only. The weak-field
equation of motion (\ref{eqmo2}) is therefore effectively two equations,
namely
\begin{equation}
X_{(0)}^{a},_{\tau \tau }-X_{(0)}^{a},_{\sigma \sigma }=0  \label{peqmo0}
\end{equation}
and
\begin{equation}
\delta X^{a},_{\tau \tau }-\delta X^{a},_{\sigma \sigma }=\alpha ^{a}
\label{peqmo1}
\end{equation}
with $\delta X^{a}\equiv \mu X_{(1)}^{a}$.

Equation (\ref{eqmo2}) was generated by first fixing a gauge and then
inserting the series expansions (\ref{expand}) into the strong-field
equation of motion (\ref{eqmo}). An alternative method of derivation is to
first write down the perturbative equation of motion in a gauge-independent
form, and then fix the gauge. A suitable gauge-independent formalism has
been developed by Battye and Carter \cite{Cart1}, and in Appendix B it is
shown that this second approach does indeed lead to the weak-field equations
of motion (\ref{peqmo0}) and (\ref{peqmo1}) in the standard gauge.

The structure of the paper from this point onwards is (I hope) very
straightforward. The unperturbed solution $X_{(0)}^{a}$ is described in
Section \ref{ACOsec} (where, however, I will write $X^{a}$ in place of $%
X_{(0)}^{a}$), the metric perturbation $h_{ab}$ is calculated in Section \ref
{pertsec}, the acceleration vector $\alpha ^{a}$ is calculated in Section
\ref{accelsec}, and the perturbed trajectory $X^{a}=X_{(0)}^{a}+\delta X^{a}$
is discussed in Section \ref{evolsec}. However, many of the details of the
calculation of $\alpha ^{a}$ are consigned to Appendix A.

\section{The Allen-Casper-Ottewill loop}

\label{ACOsec}The standard method for solving the weak-field back-reaction
problem, which was pioneered numerically by Quashnock and Spergel \cite
{Quash} over a decade ago, is to start with a loop solution $X^{a}(\tau
,\sigma )$ to the flat-space equation of motion $X^{a},_{\tau \tau
}-X^{a},_{\sigma \sigma }=0$, calculate the corresponding self-acceleration
vector $\alpha ^{a}$ at each point on the world sheet over a single period
of oscillation of the loop, integrate the resulting weak-field equation of
motion (\ref{eqmo2}) over a complete period to generate a new, perturbed
solution of the flat-space equation of motion, and then iterate the
procedure. What results is not a continuous solution to (\ref{eqmo2}) but
rather a sequence of flat-space solutions, each member of which is
constructed by debiting from the preceding member of the sequence the
gravitational energy and momentum it radiates over a single period. A
quasi-stationary approximation scheme of this type is valuable whenever the
mass per unit length $\mu $ of the string is small -- such as is the case
for GUT\ strings, which have $\mu \sim 10^{-6}$ -- as the sequence of
flat-space solutions is then almost continuous.

The Quashnock-Spergel method will be described in more detail in the next
section. In this section I will briefly explain the choice of initial loop
solution $X^{a}(\tau ,\sigma )$ to be used in the algorithm. The general
form of loop solutions to the flat-space equation of motion $X^{a},_{\tau
\tau }-X^{a},_{\sigma \sigma }=0$ is well known. In a flat background it is
always possible to choose the gauge coordinates so that $X^{0}(\tau ,\sigma
)=\tau $. The spatial components of the position function $X^{a}$ have the
form
\begin{equation}
\mathbf{X}(\tau ,\sigma )=\tfrac{1}{2}[\mathbf{a}(\tau +\sigma )+\mathbf{b}%
(\tau -\sigma )]  \label{posfunc}
\end{equation}
where for a loop in its centre-of-momentum frame the functions $\mathbf{a}$
and $\mathbf{b}$ are each periodic functions of their arguments with some
parametric period $L$. The gauge constraints $\gamma _{\tau \tau }+\gamma
_{\sigma \sigma }\equiv X_{\tau }^{2}+X_{\sigma }^{2}=0$ and $\gamma
_{\sigma \tau }\equiv X_{\sigma }\,\cdot X_{\tau }=0$ together imply that $|%
\mathbf{a}^{\prime }|^{2}=|\mathbf{b}^{\prime }|^{2}=1$, where a prime
denotes differentiation with respect to the relevant argument.

Since $\mathbf{X}(\tau +L/2,\sigma +L/2)=\mathbf{X}(\tau ,\sigma )$ for any
values of $\tau $ and $\sigma $ when $\mathbf{a}$ and $\mathbf{b}$ are
periodic functions, the fundamental oscillation period $t_{p}$ of the loop
is $L/2$ rather than the parametric period $L$. The total four-momentum of
the string loop on any surface of constant $t=\tau $ is:
\begin{equation}
p^{a}=\mu \int_{0}^{L}X_{\tau }^{a}\,d\sigma =\mu \int_{0}^{L}[1,\tfrac{1}{2}%
\mathbf{a}^{\prime }(\tau +\sigma )+\tfrac{1}{2}\mathbf{b}^{\prime }(\tau
-\sigma )]\,d\sigma =\mu L[1,\mathbf{0}]
\end{equation}
and in particular the energy of the loop (in this, the centre-of-momentum
frame) is $E=\mu L$. The corresponding angular momentum of the loop is:
\begin{equation}
\mathbf{J}=\tfrac{1}{4}\mu \int_{0}^{L}(\mathbf{a}\times \mathbf{a}^{\prime
}+\mathbf{b}\times \mathbf{b}^{\prime })\,d\sigma
\end{equation}

In the weak-field approximation the gravitational power per unit solid angle
radiated by a periodic source out to a point $\mathbf{x}$ in the wave zone,
where $r\equiv |\mathbf{x}|$ is much greater than the characteristic
dimension $L$ of the source, is given by
\begin{equation}
\frac{dP}{d\Omega }=-\frac{r^{2}}{8\pi }\sum_{j=1}^{3}n^{j}\left\langle
R^{j0}\right\rangle ,
\end{equation}
where $\mathbf{n}=\mathbf{x}/r$ and $R^{ab}$ here denotes the Ricci tensor
evaluated to second order in the metric perturbations $h_{ab}$ and to second
order in $r^{-1}$. The angled brackets indicate coarse-grained averaging,
which is performed by expanding $h_{ab}$ as a Fourier series in the
components of the field point $x^{a}$ and then eliminating all but the
zeroth-order Fourier mode from $R^{ab}$.

The resulting expression for the power radiated per unit solid angle is

\begin{equation}
\frac{dP}{d\Omega }=\frac{\omega ^{2}}{\pi }\sum_{m=1}^{\infty }m^{2}[\bar{T}%
^{ab}\bar{T}_{ab}^{*}\mathbf{-}\tfrac{1}{2}|\bar{T}_{a}^{a}\mathbf{|}^{2}]
\end{equation}
where $\omega =2\pi /t_{p}$ is the circular frequency of the source, and
\begin{equation}
\bar{T}^{ab}(m,\mathbf{n})=t_{p}^{-1}\int_{\Bbb{R}^{3}}\int_{0}^{t_{p}}%
\,T^{ab}(t,\mathbf{x})\,e^{im\omega (t-\mathbf{n\,}\cdot \,\mathbf{x}%
)}\,dt\,d^{3}x.
\end{equation}

In the case of the string stress-energy tensor (\ref{SE}), with $X^{a}=[\tau
,\frac{1}{2}(\mathbf{a}+\mathbf{b})]^{a}$ a given flat-space solution, it is
more convenient to work with the light-cone coordinates $\sigma _{\pm }=\tau
\pm \sigma $ rather than with $\tau $ and $\sigma $ themselves. The
fundamental domain $0\leq \tau <L/2$ and $0\leq \sigma <L$ is then
equivalent to $0\leq \sigma _{+}<L$ and $0\leq \sigma _{-}<L$. If $a^{\prime
c}\equiv [1,\mathbf{a}^{\prime }(\sigma _{+})]^{c}$ and $b^{\prime c}\equiv
[1,\mathbf{b}^{\prime }(\sigma _{-})]^{c}$ then $\bar{T}^{ab}=\mu
LA^{(a}B^{b)}$, where the round brackets denote symmetrisation, and
\begin{equation}
A^{c}(m,\mathbf{n})=\tfrac{1}{L}\int_{0}^{L}e^{2\pi im[\sigma _{+}-\,\mathbf{%
n\,}\cdot \,\mathbf{a}(\sigma _{+})]/L}\,a^{\prime c}(\sigma _{+})\,d\sigma
_{+}
\end{equation}
and
\begin{equation}
B^{c}(m,\mathbf{n})=\tfrac{1}{L}\int_{0}^{L}e^{2\pi im[\sigma _{-}-\,\mathbf{%
n\,}\cdot \,\mathbf{b}(\sigma _{-})]/L}\,b^{\prime c}(\sigma _{-})\,d\sigma
_{-}.
\end{equation}
In terms of the four-vectors $A^{c}$ and $B^{c}$, therefore,
\begin{equation}
\frac{dP}{d\Omega }=8\pi \mu ^{2}\sum_{m=1}^{\infty }m^{2}[(A\cdot
A^{*})(B\cdot B^{*})+|A\cdot B^{*}|^{2}-|A\cdot B|^{2}].
\end{equation}

A convenient measure of the total power radiated by a string loop is the
radiative efficiency $\gamma ^{0}\equiv \mu ^{-2}P$, where $P=\int \frac{dP}{%
d\Omega }d\Omega $ is the total power of the loop. In terms of $\gamma ^{0}$%
, the fractional energy radiated by the loop over the double oscillation
period $L=2t_{p}$ is $|\Delta E/E|\equiv (PL)/(\mu L)=\mu \gamma ^{0}$.
Numerical simulations of networks of string loops indicate that $\gamma ^{0}$
is typically of order $65$-$70$, although it may fall as low as $40$ \cite
{All-Shell, Casp-All}.

The range of $\gamma ^{0}$ has been examined analytically by Allen, Casper
and Ottewill \cite{ACO1} for a large class of string loops in which the mode
function $\mathbf{a}$ consists of two anti-parallel segments of length $%
\frac{1}{2}L$ aligned along the $z$-axis, so that
\begin{equation}
\mathbf{a}(\sigma _{+})=\left\{
\begin{array}{ll}
\sigma _{+}\,\mathbf{\hat{z}} & \text{if\qquad }0\leq \sigma _{+}<\frac{1}{2}%
L \\
(L-\sigma _{+})\,\mathbf{\hat{z}} & \text{if\qquad }\frac{1}{2}L\leq \sigma
_{+}<L
\end{array}
\right. ,  \label{a-mode}
\end{equation}
and the mode function $\mathbf{b}$ is restricted to the $x$-$y$ plane but is
otherwise unconstrained. The minimum possible value of $\gamma ^{0}$ for
loops of this class is approximately $39.0025$, and occurs when
\begin{equation}
\mathbf{b}(\sigma _{-})=\tfrac{L}{2\pi }[\cos (2\pi \sigma _{-}/L)\,\mathbf{%
\hat{x}}+\sin (2\pi \sigma _{-}/L)\,\mathbf{\hat{y}}].  \label{b-mode}
\end{equation}

The corresponding loop consists of two helical segments which are rigidly
rotating about the $z$-axis with an angular speed $\omega =4\pi /L$. The
evolution of this loop is illustrated in Figure 1, which shows the $y$-$z$
projection of the loop at times $\tau -\varepsilon =0$, $L/16$, $L/8$ and $%
3L/16$ (top row) and $\tau -\varepsilon =L/4$, $5L/16$, $3L/8$ and $7L/16$
(bottom row), where the time offset $\varepsilon $ is $0.02L$. The string
has been artificially thickened for the sake of visibility, and the $z$-axis
is also shown. The projections of the loop onto the $x$-$y$ plane are
circles of radius $L/(4\pi )$. The points at the extreme top and bottom of
the loop, where the helical segments meet and the modal tangent vector $%
\mathbf{a}^{\prime }$ is discontinuous, are technically known as kinks. They
trace out circles in the planes $z=L/4$ and $z=0$. All other points on the
loop trace out identical circles, although with varying phase lags. The net
angular momentum of the loop is $\mathbf{J}=\tfrac{1}{8\pi }\mu L^{2}\mathbf{%
\hat{z}}$.

The importance of the solution defined by the mode functions (\ref{a-mode})
and (\ref{b-mode}), which I will henceforth call \emph{the}
Allen-Casper-Ottewill (ACO) loop, lies in the fact that its radiative
efficiency $\gamma ^{0}\approx 39.0025$ is the lowest known of any loop
solution. Casper and Allen \cite{Casp-All} have examined an ensemble of
11,625 string loops, generated by numerically evolving a large set of parent
loops forward in time (in the absence of radiative effects) until only
non-intersecting child loops remained, and found that only 6 of the loops
studied had radiative efficiencies less than 42, while none at all had $%
\gamma ^{0}<40$. When examined more closely, all 6 of the low-$\gamma ^{0}$
loops were seen to have the same general shape as the ACO loop. A second
study by Casper and Allen involving another 12,830 loops yielded similar
results.

This observation has interesting implications for the evolution of a network
of string loops. If it can be shown that an ACO\ loop does not evolve
secularly towards another shape with higher radiative efficiency, then it
will be the longest-lived loop in any ensemble of loops with a given energy $%
E$ (and invariant length $L$). Furthermore, if the general trend in loop
evolution is from higher to lower radiative efficiencies -- a trend which
has not so far been demonstrated, but which would be expected if radiative
dampening acts to preferentially eliminate higher-multipole structure --
then the ACO loop would in some sense play the role of an attractor in the
ensemble of loops forming a network. However, it is conceivable that,
because the energy $E$ of a string loop falls off as $E_{0}-\mu ^{2}\gamma
^{0}t$ (since $|\Delta E|=\mu \gamma ^{0}E$ over a time period $\Delta
t=2t_{p}$, but $\Delta t=E/\mu $), any fall in the radiative efficiencies of
the loops in a network might proceed too slowly to have a significant effect
on any but the very largest loops.

What I will demonstrate below is that, in the weak-field approximation at
least, the ACO\ loop does indeed evaporate in a self-similar manner, and so $%
\gamma ^{0}$ remains constant throughout its evolution. When the weak-field
equation of motion (\ref{eqmo2}) has been integrated over a single period $%
t_{p}$, it turns out that an initial ACO\ loop with energy $E$ and invariant
length $L$ decays to form a loop with exactly the same shape, but with
smaller values $E^{\prime }=E(1-\frac{1}{2}\mu \gamma ^{0})$ and $L^{\prime
}=L(1-\frac{1}{2}\mu \gamma ^{0})$ of energy and length, plus a small
rotational phase shift.

\section{Back-reaction of the ACO\ loop}

\subsection{The metric perturbations}

\label{pertsec}The first step in the back-reaction calculation is to
evaluate the line integrals (\ref{weakh}) for the metric perturbations $%
h_{ab}$. Let $X^{a}\equiv [\tau ,\mathbf{X}]^{a}$ be a general source point
on the string loop, and $\bar{x}^{a}\equiv [\bar{t},\mathbf{\bar{x}}]^{a}$ a
general field point. The mode functions of an ACO loop with invariant length
$L$ are as given in (\ref{a-mode}) and (\ref{b-mode}), but can be simplified
slightly (by rescaling the gauge coordinates $\sigma _{+}$ and $\sigma _{-}$%
) to read
\begin{equation}
\mathbf{a}(\sigma _{+})=L|v_{+}|\,\mathbf{\hat{z}}\qquad \text{if }-\tfrac{1%
}{2}\leq v_{+}\leq \tfrac{1}{2}  \label{ACO1}
\end{equation}
and
\begin{equation}
\mathbf{b}(\sigma _{-})=\tfrac{L}{2\pi }[\cos (2\pi v_{-})\,\mathbf{\hat{x}}%
+\sin (2\pi v_{-})\,\mathbf{\hat{y}}].  \label{ACO2}
\end{equation}
where $(v_{+},v_{-})\equiv (\sigma _{+}/L,\sigma _{-}/L)$. Note that at any
point on the world sheet it is always possible to choose $v_{+}$ and $v_{-}$
so that $\tau \equiv \frac{1}{2}L(v_{+}+v_{-})$ remains equal to the
coordinate time $t$, but $v_{+}\in [-\tfrac{1}{2},\tfrac{1}{2}]$.

The spacetime coordinates (\ref{posfunc}) of the source point are then
\begin{equation}
X^{a}=\tfrac{1}{2}L[v_{+}+v_{-},\tfrac{1}{2\pi }\cos (2\pi v_{-})\,\mathbf{%
\hat{x}}+\tfrac{1}{2\pi }\sin (2\pi v_{-})\,\mathbf{\hat{y}}+|v_{+}|\,%
\mathbf{\hat{z}}]^{a}.  \label{posfunc2}
\end{equation}
For a general field point $\bar{x}^{a}$, with $\mathbf{\bar{x}}=\bar{x}\,%
\mathbf{\hat{x}}+\bar{y}\,\mathbf{\hat{y}}+\bar{z}\,\mathbf{\hat{z}}$, the
light-cone condition $\bar{t}-\tau =|\mathbf{\bar{x}}-\mathbf{X}|$ reads:
\begin{equation}
2\bar{t}/L-v_{+}-v_{-}=\{[2\bar{x}/L-\tfrac{1}{2\pi }\cos (2\pi
v_{-})]^{2}+[2\bar{y}/L-\tfrac{1}{2\pi }\sin (2\pi v_{-})]^{2}+(2\bar{z}%
/L-|v_{+}|)^{2}\}^{1/2},  \label{gamma}
\end{equation}
or equivalently
\begin{eqnarray}
&&v_{+}=-\tfrac{1}{2}(v_{-}-2\bar{t}/L-2s\bar{z}/L)  \nonumber \\
&&+\tfrac{1}{2}\{[2\bar{x}/L-\tfrac{1}{2\pi }\cos (2\pi v_{-})]^{2}+[2\bar{y}%
/L-\tfrac{1}{2\pi }\sin (2\pi v_{-})]^{2}\}/(v_{-}-2\bar{t}/L+2s\bar{z}/L)
\nonumber \\
&&  \label{gamexp}
\end{eqnarray}
where $s={}$sgn$(v_{+})$.

In particular, if the field point itself lies on the loop, and has
scale-free gauge coordinates $(\sigma _{+}/L,\sigma _{-}/L)=(u_{+},u_{-})$
so that
\begin{equation}
\lbrack \bar{t},\mathbf{\bar{x}}]=\tfrac{1}{2}L[u_{+}+u_{-},\tfrac{1}{2\pi }%
\cos (2\pi u_{-})\,\mathbf{\hat{x}}+\tfrac{1}{2\pi }\sin (2\pi u_{-})\,%
\mathbf{\hat{y}+}|u_{+}|\,\mathbf{\hat{z}}]  \label{fieldpt}
\end{equation}
then the light-cone condition (\ref{gamma}) reads:
\begin{equation}
u_{+}-v_{+}+u_{-}-v_{-}=[\tfrac{1}{2\pi ^{2}}\{1-\cos 2\pi (u_{-}-v_{-})\}+(%
\bar{s}u_{+}-sv_{+})^{2}]^{1/2}  \label{gamma2}
\end{equation}
where $\bar{s}={}$sgn$(u_{+})$. [For future convenience, the 4-vector on the
right of (\ref{fieldpt}) will abbreviated as $\bar{X}^{a}\equiv [\bar{\tau},%
\mathbf{\bar{X}}]^{a}$.]

Recall from the previous section that the ACO\ loop consists of two helical
segments bounded by kinks at $\sigma _{+}/L=-\frac{1}{2}$, $0$ and $\frac{1}{%
2}$. If the source point and field point lie on the same segment ($s\bar{s}%
=1 $) then $\bar{s}u_{+}-sv_{+}=\pm (u_{+}-v_{+})$ and (\ref{gamma2}) can be
rearranged to read
\begin{equation}
2(u_{-}-v_{-})(u_{+}-v_{+})+(u_{-}-v_{-})^{2}=\tfrac{1}{2\pi ^{2}}\{1-\cos
2\pi (u_{-}-v_{-})\}.
\end{equation}
For a fixed field point $(u_{+},u_{-})$ this equation clearly has two
possible solutions:
\begin{equation}
v_{-}=u_{-}  \label{sol1a}
\end{equation}
and
\begin{equation}
v_{+}-u_{+}=-\tfrac{1}{2}(v_{-}-u_{-})+\tfrac{1}{4\pi ^{2}}\{1-\cos 2\pi
(v_{-}-u_{-})\}/(v_{-}-u_{-}).  \label{sol1b}
\end{equation}
In view of (\ref{gamma2}) the first solution is feasible if and only if $%
v_{+}\leq u_{+}$. It is also easily verified that if the second solution
holds then $u_{+}-v_{+}+u_{-}-v_{-}\geq 0$ if and only if $v_{-}-u_{-}<0$,
in which case (\ref{sol1b}) indicates that $v_{+}>u_{+}$. So the two
solutions are mutually exclusive and fully cover the case where the source
point and field point lie on the same segment.

If on the other hand the two points lie on different segments, then $s\bar{s}%
=-1$ and $\bar{s}u_{+}-sv_{+}=\pm (u_{+}+v_{+})$. So (\ref{gamma2}) expands
to give:
\begin{equation}
(2u_{+}-v_{-}+u_{-})(-2v_{+}-v_{-}+u_{-})=\tfrac{1}{2\pi ^{2}}\{1-\cos 2\pi
(u_{-}-v_{-})\},  \label{interm}
\end{equation}
or equivalently
\begin{equation}
v_{+}=-\tfrac{1}{2}(v_{-}-u_{-})+\tfrac{1}{4\pi ^{2}}\{1-\cos 2\pi
(v_{-}-u_{-})\}/(v_{-}-u_{-}-2u_{+}).  \label{sol2}
\end{equation}
For fixed values of $u_{+}$ and $u_{-}$ this equation typically defines two
branches in the $v_{+}$-$v_{-}$ plane, one with $v_{-}-u_{-}-2u_{+}<0$ and
the other with $v_{-}-u_{-}-2u_{+}>0$. From (\ref{interm}), the combinations
$2u_{+}-v_{-}+u_{-}$ and $-2v_{+}-v_{-}+u_{-}$ are either both positive or
both negative, and since their sum $2(u_{+}-v_{+}+u_{-}-v_{-})$ must be
positive or zero from (\ref{gamma2}), the branch with $v_{-}-u_{-}-2u_{+}<0$
is the only feasible one.

Figure 2 shows the backwards light cones of two field points on the same
spacelike section $\bar{\tau}=0$ of the loop. The field point marked $A$
coincides with the kink at $u_{+}=u_{-}=0$, and its past light cone consists
of the straight line segment $v_{-}=0$ (solution (\ref{sol1a})) on the left,
and the curve $v_{+}=-\tfrac{1}{2}v_{-}+\tfrac{1}{4\pi ^{2}}(1-\cos 2\pi
v_{-})/v_{-}$ (which can interchangeably be regarded as solution (\ref{sol1b}%
) or solution (\ref{sol2})) on the right. The point marked $B$ is a general
field point, and its past light cone contains all three components
identified above: solution (\ref{sol2}) on the left, followed by the
straight-line segment (\ref{sol1a}), and then solution (\ref{sol1b}) on the
right. For future reference these three segments will be designated $\Gamma
^{-}$, $\Gamma ^{0}$ and $\Gamma ^{+}$ respectively.

It is evident from the foregoing discussion that the integration contour $%
\Gamma $ is a piecewise-smooth spacelike or null curve that wraps once
around the world sheet $\mathbf{T}$ of the loop. When calculating the metric
perturbation $h_{ab}$ or its spacetime derivatives $h_{ab},_{c}$ at a given
field point $[\bar{t},\mathbf{\bar{x}}]$ it is possible in principle to use $%
\sigma $ or $v_{+}$ or $v_{-}$ as the integration variable on $\Gamma $, as
circumstances warrant. From (\ref{lightcon}) and the fact that $\sigma =%
\frac{1}{2}L(v_{+}-v_{-})$ and $\bar{t}-\tau =|\mathbf{\bar{x}}-\mathbf{X}|$%
, the relevant transformation rules are:
\begin{equation}
|\mathbf{\bar{x}}-\mathbf{X}|^{-1}|1-\mathbf{n}\cdot \mathbf{X}_{\tau
}|^{-1}d\sigma =L\Delta _{-}^{-1}dv_{+}=-L\Delta _{+}^{-1}dv_{-}
\label{transf}
\end{equation}
where
\begin{equation}
\Delta _{\pm }\equiv \bar{t}-\tau -2(\mathbf{\bar{x}}-\mathbf{X})\cdot
\mathbf{X}_{\pm }
\end{equation}
and it is understood that $v_{+}$ is varied with $v_{-}$ held constant, and
vice versa. Because an explicit formula (\ref{gamexp}) exists for $v_{+}$ as
a function of $v_{-}$ on $\Gamma $, the most useful choice of integration
variable will normally be $v_{-}$, in which case (\ref{weakh2}) becomes
\begin{equation}
h_{ab}(\bar{t},\mathbf{\bar{x}})=4\mu L\int_{\Gamma }\Delta _{+}^{-1}\Psi
_{ab}\,dv_{-}  \label{weakh3}
\end{equation}

Now, in terms of the derivatives of the light-cone coordinates $\sigma _{+}$
and $\sigma _{-}$ the formula (\ref{source}) for the source function $\Psi
^{ab}$ reads:
\begin{equation}
\Psi ^{ab}=2(X_{+}^{a}\,X_{-}^{b}+X_{-}^{a}\,X_{+}^{b})-2\,\eta
^{ab}\,X_{+}\cdot X_{-}
\end{equation}
where $X_{\pm }^{a}=[\tfrac{1}{2},\mathbf{X}_{\pm }]^{a}$, and $\mathbf{X}%
_{+}=\frac{1}{2}s\,\mathbf{\hat{z}}$ and $\mathbf{X}_{-}=\frac{1}{2}(-\sin
2\pi v_{-}\,\mathbf{\hat{x}}+\cos 2\pi v_{-}\,\mathbf{\hat{y}})$ are
orthogonal vectors. Hence, $X_{+}\cdot X_{-}=\frac{1}{4}$ and
\begin{equation}
\Psi _{ab}(s,v_{-})=\frac{1}{2}\left[
\begin{array}{cccc}
1 & \sin 2\pi v_{-} & -\cos 2\pi v_{-} & -s \\
\sin 2\pi v_{-} & 1 & 0 & -s\sin 2\pi v_{-} \\
-\cos 2\pi v_{-} & 0 & 1 & s\cos 2\pi v_{-} \\
-s & -s\sin 2\pi v_{-} & s\cos 2\pi v_{-} & 1
\end{array}
\right] _{ab}.  \label{source2}
\end{equation}

Also, since
\begin{equation}
-L\Delta _{+}^{-1}=2(v_{-}-2\bar{t}/L+2s\bar{z}/L)^{-1}
\end{equation}
the integral (\ref{weakh3}) for $h_{ab}$ reads
\begin{equation}
h_{ab}(\bar{t},\mathbf{\bar{x}})=-8\mu \int_{\Gamma }(v_{-}-2\bar{t}/L+2s%
\bar{z}/L)^{-1}\Psi _{ab}\,dv_{-}
\end{equation}
where the limits on $v_{-}$ are fixed by the requirement that $v_{+}$ range
from $-\frac{1}{2}$ to $0$ (with $s=-1$) and then from $0$ to $\frac{1}{2}$
(with $s=1$). Note here that in view of (\ref{transf}) and the fact that $%
\Delta _{+}$ and $\Delta _{-}$ are non-negative functions, $v_{-}$ is a
non-increasing function of $v_{+}$ on $\Gamma $.

If $V_{-1}$, $V_{0}$ and $V_{1}$ denote the values of $v_{-}$ when $v_{+}=-%
\frac{1}{2}$, $0$ and $\frac{1}{2}$ respectively, then the equation (\ref
{gamma}) for $\Gamma $ indicates that
\begin{equation}
2\bar{t}/L-V_{k}-\tfrac{k}{2}=\{[2\bar{x}/L-\tfrac{1}{2\pi }\cos (2\pi
V_{k})]^{2}+[2\bar{y}/L-\tfrac{1}{2\pi }\sin (2\pi V_{k})]^{2}+(2\bar{z}/L-|%
\tfrac{k}{2}|)^{2}\}^{1/2}  \label{roots}
\end{equation}
in each of the three cases $k=-1$, $0$ and $1$. In particular, it is evident
that if $V_{1}$ satisfies (\ref{roots}) with $k=1$ then $V_{-1}=V_{1}+1$
satisfies (\ref{roots}) when $k=-1$. So $V_{1}\leq V_{0}\leq V_{-1}\equiv
V_{1}+1$, and (\ref{weakh3}) becomes
\begin{equation}
h_{ab}(\bar{t},\mathbf{\bar{x}})=8\mu \int_{V_{1}}^{V_{0}}(v_{-}-\psi
_{-})^{-1}\Psi _{ab}|_{s=1}\,dv_{-}+8\mu \int_{V_{0}}^{V_{1}+1}(v_{-}-\psi
_{+})^{-1}\Psi _{ab}|_{s=-1}\,dv_{-}  \label{metpert}
\end{equation}
where $\psi _{\pm }=2\bar{t}/L\pm 2\bar{z}/L$.

In the case of greatest interest, where the field point $[\bar{t},\mathbf{%
\bar{x}}]$ is a point (\ref{fieldpt}) on the loop, if $u_{+}>0$ then $%
V_{0}=u_{-}$ and (from (\ref{gamma2}))
\begin{equation}
V_{1}=u_{-}+W_{1}(u_{+})\qquad \text{where\qquad }u_{+}=\tfrac{1}{2}+\tfrac{1%
}{2}W_{1}-\tfrac{1}{4\pi ^{2}}(1-\cos 2\pi W_{1})/W_{1}  \label{W1}
\end{equation}
while if $u_{+}<0$ then $V_{1}=u_{-}-1$ and
\begin{equation}
V_{0}=u_{-}+W_{0}(u_{+})\qquad \text{where\qquad }u_{+}=\tfrac{1}{2}W_{0}-%
\tfrac{1}{4\pi ^{2}}(1-\cos 2\pi W_{0})/W_{0}.  \label{W0}
\end{equation}
Note that $W_{1}\in (-1,0)$ for $u_{+}\in (0,\frac{1}{2})$ and $W_{0}\in
(-1,0)$ for $u_{+}\in (-\frac{1}{2},0)$. In fact, $W_{1}(u_{+})=W_{0}(u_{+}-%
\frac{1}{2})$ for all $u_{+}\in (0,\frac{1}{2})$. At each of the kink points
$u_{+}=-\frac{1}{2}$, $0$ and $\frac{1}{2}$ the values of $(V_{1},V_{0})$
tend to $(u_{-},u_{-})$ as $u_{+}$ approaches the kink value from below, and
tend to $(u_{-}-1,u_{-})$ as $u_{+}$ approaches the kink value from above.

Collecting together the results of the preceding paragraphs gives the
formula
\begin{equation}
h_{ab}(\bar{t},\mathbf{\bar{x}})=8\mu \int_{V_{1}-\psi _{-}}^{V_{0}-\psi
_{-}}w^{-1}\Psi _{ab}(1,w+\psi _{-})\,dw+8\mu \int_{V_{0}-\psi
_{+}}^{V_{1}+1-\psi _{+}}w^{-1}\Psi _{ab}(-1,w+\psi _{+})\,dw  \label{weakh4}
\end{equation}
where the components (\ref{source2}) of $\Psi _{ab}$ are all smooth, and all
except $\Psi _{xy}$ and $\Psi _{yx}$ are non-zero. The two integrals in (\ref
{weakh4}) are guaranteed to converge if the field point $[\bar{t},\mathbf{%
\bar{x}}]$ lies off the string, but not if it lies on the string. In the
second case, $\psi _{\pm }=u_{-}+u_{+}\pm |u_{+}|$ and so $V_{0}-\psi _{-}=0$
if $u_{+}>0$ while $V_{1}+1-\psi _{+}=0$ if $u_{+}<0$. Thus at least one of
the integrals in (\ref{weakh4}) diverges.

To investigate more fully the behaviour of the metric perturbations near the
world sheet $\mathbf{T}$, suppose that
\begin{equation}
\lbrack \bar{t},\mathbf{\bar{x}}]=[\bar{\tau},\mathbf{\bar{X}}]+[\delta \bar{%
t},\delta \mathbf{\bar{x}}]
\end{equation}
where $\delta \bar{t}$ and $\delta \mathbf{\bar{x}}$ are assumed to be
small. Then if $u_{+}>0$ the equation (\ref{roots}) with $k=0$ can be solved
for $V_{0}$ to second order in $[\delta \bar{t},\delta \mathbf{\bar{x}}]$ to
give
\begin{equation}
V_{0}-\psi _{-}\approx -\tfrac{2}{u_{+}L^{2}}\{\delta \bar{x}^{2}+\delta
\bar{y}^{2}+(\delta \bar{t}-\delta \bar{z})^{2}+2(\delta \bar{t}-\delta \bar{%
z})[\delta \bar{x}\sin (2\pi u_{-})-\delta \bar{y}\cos (2\pi u_{-})]\}
\label{div1}
\end{equation}
whereas if $u_{+}<0$ the equation (\ref{roots}) with $k=1$ can be solved for
$V_{1}$ to give
\begin{equation}
V_{1}+1-\psi _{+}\approx -\tfrac{2}{(u_{+}+\frac{1}{2})L^{2}}\{\delta \bar{x}%
^{2}+\delta \bar{y}^{2}+(\delta \bar{t}+\delta \bar{z})^{2}+2(\delta \bar{t}%
+\delta \bar{z})[\delta \bar{x}\sin (2\pi u_{-})-\delta \bar{y}\cos (2\pi
u_{-})]\}.  \label{div2}
\end{equation}

Here, the quadratic forms
\begin{eqnarray}
Q_{\pm } &=&\delta \bar{x}^{2}+\delta \bar{y}^{2}+(\delta \bar{t}\pm \delta
\bar{z})^{2}+2(\delta \bar{t}\pm \delta \bar{z})[\delta \bar{x}\sin (2\pi
u_{-})-\delta \bar{y}\cos (2\pi u_{-})]  \nonumber \\
&\equiv &2\Psi _{ab}(\mp 1,u_{-})[\delta \bar{t},\delta \mathbf{\bar{x}}%
]^{a}[\delta \bar{t},\delta \mathbf{\bar{x}}]^{b}
\end{eqnarray}
are positive semi-definite and vanish on the 2-surface in $[\delta \bar{t}%
,\delta \mathbf{\bar{x}}]$ space spanned by the vectors $[1,\mp \mathbf{\hat{%
z}}]$ and $[1,-\sin (2\pi u_{-})\mathbf{\hat{x}}+\cos (2\pi u_{-})\mathbf{%
\hat{y}}]$, which is just the tangent plane to the string at the point $[%
\bar{\tau},\mathbf{\bar{X}}]$. Thus, if the field point $[\bar{t},\mathbf{%
\bar{x}}]$ approaches a point on the string from any direction outside the
tangent plane, the divergent contribution to $h_{ab}$ has the form
\begin{equation}
h_{ab}\approx 8\mu \Psi _{ab}(\pm 1,\psi _{\mp })\ln Q_{\mp }  \label{h-div}
\end{equation}
where the upper sign applies when $u_{+}>0$ and the lower sign when $u_{+}<0$%
. Note also from (\ref{div1}) and (\ref{div2}) that the divergence is more
complicated at the kink points $u_{+}=0$ and $u_{+}=-\frac{1}{2}$.

The divergence of the metric perturbation $h_{ab}$ on the world sheet is to
be expected (see for example \cite{Vile-Shell}, p. 216), and is simply a
reflection of the conical singularity that is present at every ordinary
point on a zero-thickness cosmic string. For example, the exact metric about
a static straight string lying along the $z$-axis has the ``isotropic'' form
\cite{Linet}
\begin{equation}
ds^{2}=dt^{2}-dz^{2}-\rho ^{-8\mu }(dx^{2}+dy^{2})
\end{equation}
where $\rho =(x^{2}+y^{2})^{1/2}$. In the limit of small $\mu $ the only
non-zero metric perturbations are $h_{xx}=h_{yy}=8\mu \ln \rho $, which
satisfy the Lorentz gauge conditions $h_{a}^{b},_{b}=\frac{1}{2}h,_{a}$ but
also diverge logarithmically as $\rho \rightarrow 0$.

Two quantities that will play an important role in Section \ref{evolsec} in
fixing the initial data for the perturbation $\delta X^{a}$ are the values
of the functions $h_{ab}\bar{X}_{+}^{a}\bar{X}_{+}^{b}$ and $h_{ab}\bar{X}%
_{-}^{a}\bar{X}_{-}^{b}$ on the world sheet $\mathbf{T}$. Since $\Psi
_{ab}X_{+}^{b}=\Psi _{ab}X_{-}^{b}=0$ at all ordinary points on $\mathbf{T}$%
, the leading-order divergence (\ref{h-div}) in $h_{ab}$ makes no
contribution to these functions.

At the next highest order in $[\delta \bar{t},\delta \mathbf{\bar{x}}]$, the
metric perturbation has the form
\begin{equation}
h_{ab}=4\mu \left[
\begin{array}{cccc}
I_{+}+I_{-} & S_{+}+S_{-} & -C_{+}-C_{-} & -I_{+}+I_{-} \\
S_{+}+S_{-} & I_{+}+I_{-} & 0 & -S_{+}+S_{-} \\
-C_{+}-C_{-} & 0 & I_{+}+I_{-} & C_{+}-C_{-} \\
-I_{+}+I_{-} & -S_{+}+S_{-} & C_{+}-C_{-} & I_{+}+I_{-}
\end{array}
\right] _{ab}
\end{equation}
where, for $0<u_{+}<\frac{1}{2}$,
\begin{equation}
I_{+}=-\ln |W_{1}|\text{,\qquad }I_{-}=\ln D_{+}-\ln 2u_{+}
\end{equation}
\begin{eqnarray}
(S_{+},C_{+}) &=&(\cos 2\pi u_{-},-\sin 2\pi u_{-})\func{Si}(2\pi
|W_{1}|)
\nonumber \\
&&-(\sin 2\pi u_{-},\cos 2\pi u_{-})[\func{Ci}(2\pi |W_{1}|)-\gamma _{\text{E%
}}-\ln 2\pi ]\allowbreak
\end{eqnarray}
and
\begin{eqnarray}
(S_{-},C_{-}) &=&(-\cos 2\pi (2u_{+}+u_{-}),\sin 2\pi (2u_{+}+u_{-}))[\func{%
Si}(2\pi D_{+})-\func{Si}(4\pi u_{+})]  \nonumber \\
&&+(\sin 2\pi (2u_{+}+u_{-}),\cos 2\pi (2u_{+}+u_{-}))[\func{Ci}(2\pi D_{+})-%
\func{Ci}(4\pi u_{+})]  \nonumber \\
&&
\end{eqnarray}
with $\func{Si}(x)=\int_{0}^{x}w^{-1}\sin w\,dw$ and $\func{Ci}%
(x)=-\int_{x}^{\infty }w^{-1}\cos w\,dw$ the usual sine and cosine
integrals, and $\gamma _{\text{E}}\approx 0.5772$ Euler's constant. (In
fact, $\func{Ci}(2\pi x)-\gamma _{\text{E}}-\ln 2\pi =\ln
x-\int_{0}^{x}x^{-1}(1-\cos 2\pi x)dx$.) Also, the function $D_{+}(u_{+})$
is defined by
\begin{equation}
2\pi D_{+}=2\pi (2u_{+}-W_{1}-1)\equiv -\tfrac{1}{\pi }(1-\cos 2\pi
W_{1})/W_{1}
\end{equation}
while $W_{1}$ is defined implicitly as a function of $u_{+}$ through (\ref
{W1}).

Substituting $\bar{X}_{+}^{a}=\frac{1}{2}[1,\bar{s}\,\mathbf{\hat{z}}]^{a}$
and $\bar{X}_{-}^{a}=\frac{1}{2}[1,-\sin 2\pi u_{-}\,\mathbf{\hat{x}}+\cos
2\pi u_{-}\,\mathbf{\hat{y}}]^{a}$ gives
\begin{equation}
h_{ab}\bar{X}_{+}^{a}\bar{X}_{+}^{b}=4\mu I_{-}=4\mu (\ln D_{+}-\ln 2u_{+})
\label{gpert1}
\end{equation}
and
\begin{eqnarray}
h_{ab}\bar{X}_{-}^{a}\bar{X}_{-}^{b} &=&2\mu (I_{+}+I_{-})-2\mu
[(S_{+}+S_{-})\sin 2\pi u_{-}+(C_{+}+C_{-})\cos 2\pi u_{-}]  \nonumber \\
&=&2\mu (\ln D_{+}-\ln 2u_{+}-\ln |W_{1}|)+2\mu [\func{Ci}(2\pi
|W_{1}|)-\gamma _{\text{E}}-\ln 2\pi ]  \nonumber \\
&&-2\mu [\func{Si}(2\pi D_{+})-\func{Si}(4\pi u_{+})]\allowbreak \sin (4\pi
u_{+})  \nonumber \\
&&-2\mu [\func{Ci}(2\pi D_{+})-\func{Ci}(4\pi u_{+})]\allowbreak \cos (4\pi
u_{+})  \label{gpert2}
\end{eqnarray}
when $0<u_{+}<\frac{1}{2}$. The function $h_{ab}\bar{X}_{+}^{a}\bar{X}%
_{+}^{b}$ diverges as $4\mu \ln u_{+}$ as $u_{+}\rightarrow 0$, and as $%
\frac{4}{3}\mu \ln (1-2u_{+})$ as $u_{+}\rightarrow \frac{1}{2}$. However,
the function $h_{ab}\bar{X}_{-}^{a}\bar{X}_{-}^{b}$ is bounded, and
converges to $2\mu [\func{Ci}(2\pi )-\gamma _{\text{E}}-\ln 2\pi ]\approx
-4.\,\allowbreak 875\,3\mu $ at both $u_{+}=0$ and $u_{+}=\frac{1}{2}$.

In the case where $-\frac{1}{2}<u_{+}<0$, it turns out that $%
I_{+}(u_{+})=I_{-}(u_{+}+\frac{1}{2})$ and $I_{-}(u_{+})=I_{+}(u_{+}+\frac{1%
}{2})$, and analogous identities hold for $S_{\pm }$ and $C_{\pm }$. The
functions $h_{ab}\bar{X}_{+}^{a}\bar{X}_{+}^{b}$ and $h_{ab}\bar{X}_{-}^{a}%
\bar{X}_{-}^{b}$ are therefore periodic functions of $u_{+}$ with period $%
\frac{1}{2}$.

A similar calculation of the leading-order contributions to $h_{ab}$ near
the kink point $u_{+}=0$ gives:
\begin{eqnarray}
(I_{+},S_{+},C_{+}) &\approx &(1,\sin 2\pi u_{-},\cos 2\pi u_{-})\ln
(L^{-1}Q_{-}/Q_{0})  \nonumber \\
&&+(0,\cos 2\pi u_{-},-\sin 2\pi u_{-})\func{Si}(2\pi )  \nonumber \\
&&-(0,\sin 2\pi u_{-},\cos 2\pi u_{-})[\func{Ci}(2\pi )-\gamma _{\text{E}%
}-\ln 2\pi ]  \label{kinkplus}
\end{eqnarray}
and
\begin{equation}
(I_{-},S_{-},C_{-})\approx (1,\sin 2\pi u_{-},\cos 2\pi u_{-})\ln
(4L^{-1}Q_{0})  \label{kinkminus}
\end{equation}
where
\begin{equation}
Q_{0}=\delta \bar{t}+\delta \bar{x}\sin 2\pi u_{-}-\delta \bar{y}\cos 2\pi
u_{-}\equiv 2\eta _{ab}\bar{X}_{-}^{a}[\delta \bar{t},\delta \mathbf{\bar{x}}%
]^{b}.
\end{equation}
Hence, $h_{ab}\bar{X}_{-}^{a}\bar{X}_{-}^{b}=$ $2\mu [\func{Ci}(2\pi
)-\gamma _{\text{E}}-\ln 2\pi ]$ at the kink point and the function $h_{ab}%
\bar{X}_{-}^{a}\bar{X}_{-}^{b}$ is continuous there. By contrast, $\bar{X}%
_{+}^{a}$ has a jump discontinuity at the kink point, and $h_{ab}\bar{X}%
_{+}^{a}\bar{X}_{+}^{b}$ (as well as containing formal logarithmic
divergences) is undefined even for non-zero $[\delta \bar{t},\delta \mathbf{%
\bar{x}}]$.

\subsection{The self-acceleration vector}

\label{accelsec}Although $h_{ab}$ diverges at all points on the loop, the
self-acceleration vector $\alpha ^{a}$ defined in (\ref{accel}) typically
does not. This result was first established by Quashnock and Spergel \cite
{Quash} for a generic point on a smooth loop, but does not extend to
non-generic points such as cusps and kinks.\footnote{%
It should be emphasised that the Quashnock-Spergel result holds only to
linear order in the expansion parameter $\mu $. Copeland, Haws and Hindmarsh
\cite{CHH} have made the claim that the integral for the self-acceleration
vector $\alpha ^{a}$ contains a local divergence of the form $\mu ^{2}\ln
\rho $, where $\rho $ is the normal distance from the world sheet. Carter
and Battye \cite{Cart2} have more recently disputed this claim, maintaining
that the divergence is cancelled by a counter-term that was previously
neglected. Since neither analysis is based on a full second-order expansion
of the equations of motion ((\ref{eqmo}) and the equation $K^{c}=0$ of
Appendix B in the respective cases), this dispute remains unresolved.
However, the question has no relevance for the purely first-order treatment
of the back-reaction problem offered here.} In terms of the light-cone
derivatives $\bar{X}_{+}^{a}$ and $\bar{X}_{-}^{a}$ of $\bar{X}^{a}=[\bar{%
\tau},\mathbf{\bar{X}}]^{a}$, the expression (\ref{accel}) for $\alpha ^{a}$
at a point $(u_{+},u_{-})$ on the string loop can be rewritten as
\begin{equation}
\alpha ^{a}(u_{+},u_{-})=-\eta ^{ad}(2h_{db},_{c}-h_{bc},_{d})(\bar{X}%
_{+}^{b}\,\bar{X}_{-}^{c}+\bar{X}_{-}^{b}\,\bar{X}_{+}^{c})  \label{accel2}
\end{equation}
where it is understood that the right-hand side is evaluated by taking the
limit $[\bar{t},\mathbf{\bar{x}}]\rightarrow [\bar{\tau},\mathbf{\bar{X}}]$,
from any direction \emph{outside} the tangent plane, in the metric
derivatives $h_{ab},_{c}$.

Now the ACO\ loop, as described by (\ref{ACO1}) and (\ref{ACO2}), rotates
uniformly about the $z$-axis, and so the self-acceleration vector $\alpha
^{a}$ is uniformly rotating as well. In order to simplify the calculation of
$\alpha ^{a}$, it is useful to premultiply $\alpha ^{a}$ by the matrix
\begin{equation}
M_{b}^{a}=\left[
\begin{array}{llll}
1 & 0 & 0 & 0 \\
0 & \cos 2\pi u_{-} & \sin 2\pi u_{-} & 0 \\
0 & -\sin 2\pi u_{-} & \cos 2\pi u_{-} & 0 \\
0 & 0 & 0 & 1
\end{array}
\right] _{ab}
\end{equation}
the effect of which is to map each horizontal radius vector joining the
rotation axis to a point on the loop into a vector parallel to the $x$-axis.
In other words, $\alpha ^{L}\equiv M_{b}^{x}\alpha ^{b}$ is the lateral
component of the loop's horizontal acceleration and $\alpha ^{N}\equiv
M_{b}^{y}\alpha ^{b}$ is its normal component.

To demonstrate that $\alpha ^{a}$ is well defined away from the kink points
at $u_{+}=0$ and $\pm \frac{1}{2}$, and to explicitly calculate the
components of the co-rotating acceleration vector $M_{b}^{a}\alpha ^{b}$, is
straightforward but tedious. The details can be found in Appendix A. It
turns out that $\alpha ^{t}$, $\alpha ^{L}$ and $\alpha ^{N}$ are periodic
functions of $u_{+}$, and $\alpha ^{z}$ is an anti-periodic function, with
period $\frac{1}{2}$.

For $0<u_{+}<\frac{1}{2}$, the components of $M_{b}^{a}\alpha ^{b}$ are
\begin{equation}
\alpha ^{t}=\alpha ^{z}=32\pi ^{2}\mu L^{-1}F(W_{1})W_{1}(1-\cos 2\pi W_{1})
\end{equation}
and
\begin{eqnarray}
&&\left[
\begin{array}{l}
\alpha ^{L} \\
\alpha ^{N}
\end{array}
\right] =-32\pi \mu L^{-1}  \nonumber \\
&&\times \left[
\begin{array}{l}
\{\func{Si}(4\pi u_{+})-\func{Si}(2\pi D_{+})\}\sin 4\pi u_{+}+\{\func{Ci}%
(4\pi u_{+})-\func{Ci}(2\pi D_{+})\}\cos 4\pi u_{+} \\
-\{\func{Si}(4\pi u_{+})-\func{Si}(2\pi D_{+})\}\cos 4\pi u_{+}+\{\func{Ci}%
(4\pi u_{+})-\func{Ci}(2\pi D_{+})\}\sin 4\pi u_{+}
\end{array}
\right]  \nonumber \\
&&+32\pi ^{2}\mu L^{-1}F(W_{1})W_{1}\left[
\begin{array}{l}
2\pi W_{1}\cos 2\pi W_{1}-\sin 2\pi W\allowbreak _{1} \\
\cos 2\pi W_{1}-1+2\pi W_{1}\sin 2\pi W_{1}
\end{array}
\right]
\end{eqnarray}
where
\begin{equation}
F(W_{1})=(2\pi ^{2}W_{1}^{2}+1-\cos 2\pi W_{1}-2\pi W_{1}\sin 2\pi
W_{1})^{-1},
\end{equation}
\begin{equation}
2\pi D_{+}=-\tfrac{1}{\pi }(1-\cos 2\pi W_{1})/W_{1}.
\end{equation}
as before, and $W_{1}$ is defined implicitly as a function of $u_{+}$
through
\begin{equation}
4\pi u_{+}=2\pi (1+W_{1})-\tfrac{1}{\pi }(1-\cos 2\pi W_{1})/W_{1}
\end{equation}
(see (\ref{W1})).

The dependence of the acceleration component $\alpha ^{t}$ on $u_{+}$ is
shown in Figure 3. The interval $u_{+}\in (0,\frac{1}{2})$ corresponds to $%
W_{1}\in (-1,0)$. Since $\alpha ^{t}$ diverges as $32\mu L^{-1}W_{1}^{-1}$
near $W_{1}=0$ while $u_{+}-\frac{1}{2}\approx \frac{1}{6}\pi ^{2}W_{1}^{3}$
for small values of $W_{1}<0$, the singularity in $\alpha ^{t}$ at $u_{+}=0$
has $\alpha ^{t}\approx -32(\frac{1}{6}\pi ^{2})^{1/3}\mu
L^{-1}/|u_{+}|^{1/3}$ for $u_{+}\lesssim 0$. Note however that the
singularity in $\alpha ^{t}$ is one-sided, as $\alpha ^{t}\approx -32\pi
^{2}\mu L^{-1}(1+W_{1})^{2}$ for $W_{1}$ near $-1$, while $u_{+}\approx
\frac{1}{2}(1+W_{1})$, and so $\alpha ^{t}\approx -128\pi ^{2}\mu
L^{-1}u_{+}^{2}$ for $u_{+}\gtrsim 0$. Similar remarks apply to $\alpha ^{z}$%
, save that $\alpha ^{z}>0$ for $u_{+}\in (-\frac{1}{2},0)$ and $\alpha
^{z}<0$ for $u_{+}\in (0,\frac{1}{2})$.

The point $u_{+}\equiv \bar{\tau}+\bar{\sigma}=0$ of course marks the
location of the lower of the two kinks on the string, which propagates
around the string in the direction of increasing $\bar{\sigma}$. For any
fixed value of $\bar{\tau}$, points with $u_{+}\gtrsim 0$ lie ahead of the
kink and have yet to be disturbed by its passing, whereas points with $%
u_{+}\lesssim 0$ lie just behind the kink and so have recently been subject
to severe gravitational stresses. This is the reason for the one-sided
nature of the singularity in $\alpha ^{t}$ and $\alpha ^{z}$ at $u_{+}=0$.
Also, $\alpha ^{t}\leq 0$ because the string is radiating energy, whereas $%
\alpha ^{z}$ has different signs on the two branches of the loop because it
is directed counter to the $z$-component of the string's 4-velocity $X_{\tau
}^{a}$, which is negative for $u_{+}\in (-\frac{1}{2},0)$ and positive for $%
u_{+}\in (0,\frac{1}{2})$.

The two remaining components of the acceleration vector, $\alpha ^{L}$ and $%
\alpha ^{N}$, are plotted in Figures 4 and 5 respectively. Both are
divergent at $u_{+}=0$, with $\alpha ^{L}\approx \frac{32}{3}\pi \mu
L^{-1}\ln |u_{+}|$ for $u_{+}\lesssim 0$ and $\alpha ^{L}\approx 32\pi \mu
L^{-1}\ln u_{+}$ for $u_{+}\gtrsim 0$, while $\alpha ^{N}\approx -32(\frac{1%
}{6}\pi ^{2})^{1/3}\mu L^{-1}/|u_{+}|^{1/3}$ for $u_{+}\lesssim 0$ and $%
\alpha ^{N}\approx 128\pi ^{2}\mu L^{-1}u_{+}\ln u_{+}$ for $u_{+}\gtrsim 0$%
. The normal component $\alpha ^{N}$ is negative almost everywhere because
it acts to torque down the rotational motion of the loop, and like $\alpha
^{t}$ and $\alpha ^{z}$ it is strongest in the immediate wake of the two
kinks. The lateral component $\alpha ^{L}$ is also negative, but its
singularity at $u_{+}=0$ is more symmetric. The lateral acceleration is
presumably a response to the imbalance between the string tension and the
centripetal force caused by the rotational torquing. Its net effect is to
induce lateral shrinkage of the loop as it radiates.

Although all four components of the acceleration vector $\alpha ^{a}$ are
divergent at the kink points $u_{+}=-\frac{1}{2}$, $0$ and $\frac{1}{2}$,
the singularities are all integrable, in the sense that $\int \alpha
^{a}du_{+}$ exists on any closed sub-interval of $[-\frac{1}{2},\frac{1}{2}]$%
. This is turn means that the weak-field equations of motion for the ACO
loop are integrable as well, as will be seen shortly.

It could be argued that the method adopted here, of evaluating $\alpha ^{a}$
away from the kink points then integrating the equations of motion through
the kinks, ignores the possible presence of other localised (delta-function
or non-integrable) singularities in $\alpha ^{a}$ at the kink points
themselves. There seems to be no way of ruling out this possibility short of
solving the full set of field equations for a finite-thickness string and
taking the limit as the thickness goes to zero. However, it should be noted
that a formal calculation of $\alpha ^{a}$ at the kink point $u_{+}=0$ using
only the logarithmically divergent contributions (\ref{kinkplus}) and (\ref
{kinkminus}) to $h_{ab}$ gives $\alpha ^{a}=0$ irrespective of the value
chosen for $\bar{s}$ in the tangent vector $\bar{X}_{+}^{a}=\frac{1}{2}[1,%
\bar{s}\,\mathbf{\hat{z}}]^{a}$. (This remark should not be taken to mean
that $\alpha ^{a}$ actually vanishes at the kink points, as terms of order $%
[\delta \bar{t},\delta \mathbf{\bar{x}}]$ in $h_{ab}$ contribute additional
terms to $\alpha ^{a}$ which have no well-defined limits when $[\delta \bar{t%
},\delta \mathbf{\bar{x}}]\rightarrow 0$.\footnote{%
More explicitly, an expansion of $h_{ab}$ about the kink point $u_{+}=0$
reads, to linear order in $\delta \bar{x}^{a}=[\delta \bar{t},\delta \mathbf{%
\bar{x}}]^{a}$,
\begin{eqnarray*}
h_{ab} &=&8\mu [\Psi _{ab}(1,u_{-})\ln (L^{-1}Q_{-}/Q_{0})+\Psi
_{ab}(-1,u_{-})\ln (4L^{-1}Q_{0})+H_{ab}(u_{-}) \\
&&+F_{ab}^{-}(u_{-})\{2(\delta \bar{t}/L-\delta \bar{z}/L)\ln
(L^{-1}Q_{-}/Q_{0})-L^{-1}Q_{-}/Q_{0}\} \\
&&+F_{ab}^{+}(u_{-})\{2(\delta \bar{t}/L+\delta \bar{z}/L)\ln
(4L^{-1}Q_{0})+L^{-1}Q_{+}/Q_{0}\}+\cdots ]
\end{eqnarray*}
where the ellipsis ($\cdots $) includes terms of first-order in $\delta \bar{%
x}^{a}$ even more complicated than those shown. The corresponding
acceleration vector $\alpha ^{a}$, calculated from (\ref{accel2}), contains
no contributions from the first two, logarithmically divergent, terms.
However, the terms proportional to $F_{ab}^{-}$ and $F_{ab}^{+}$, and the
other linear terms, do formally contribute to $\alpha ^{a}$, with
coefficients that typically depend on $\bar{s}={}$sgn$(u_{+})$. The
derivatives with respect to $\delta \bar{x}^{a}$ of all these terms are
undefined in the limit as $\delta \bar{x}^{a}\rightarrow 0$.}) Furthermore,
as will be seen, the energy loss of the loop over a parametric period
exactly matches the radiative flux calculated for the ACO\ loop in \cite
{ACO1}, so there are good reasons for believing that the acceleration vector
$\alpha ^{a}$ has been represented accurately.

\subsection{Secular evolution of the ACO loop}

\label{evolsec}Once the acceleration vector
\begin{equation}
\alpha ^{a}=[\alpha ^{t},(\alpha ^{L}\cos 2\pi u_{-}-\alpha ^{N}\sin 2\pi
u_{-})\mathbf{\hat{x}}+(\alpha ^{L}\sin 2\pi u_{-}+\alpha ^{N}\cos 2\pi
u_{-})\mathbf{\hat{y}}+\alpha ^{z}\mathbf{\hat{z}}]^{a}  \label{accvec}
\end{equation}
is known as a function of $u_{+}$ and $u_{-}$, it is a straightforward
matter to use (\ref{eqmo2}) to calculate the secular perturbations in the
trajectory of the ACO loop due to gravitational back-reaction. In terms of
the derivatives with respect to the scaled gauge coordinates $\sigma
_{+}=Lu_{+}$ and $\sigma _{-}=Lu_{-}$ the first-order equation of motion (%
\ref{peqmo1}) reads
\begin{equation}
\delta X,_{+-}^{a}=\tfrac{1}{4}\alpha ^{a}(\sigma _{+},\sigma _{-}).
\label{eqmo3}
\end{equation}

The initial data for this equation on the curve \footnote{%
The bars on $\bar{X}^{a}$, $\bar{\tau}$ and $\bar{\sigma}$ can henceforth be
dropped without risk of confusion.} $\tau \equiv \frac{1}{2}(\sigma
_{+}+\sigma _{-})=0$ are fixed by requiring $\delta X_{+}^{a}|_{\tau =0}$
and $\delta X_{-}^{a}|_{\tau =0}$ to satisfy the first-order gauge
constraint (\ref{const1}), while preserving the initial configuration $%
X^{a}|_{\tau =0}$ of the loop and the tangent space to the loop at each
point on the initial curve. This can be done by writing
\begin{equation}
\delta X_{+}^{a}|_{\tau =0}=k^{+}X_{+}^{a}|_{\tau =0}+l^{+}X_{-}^{a}|_{\tau
=0}
\end{equation}
and
\begin{equation}
\delta X_{-}^{a}|_{\tau =0}=k^{-}X_{-}^{a}|_{\tau =0}+l^{-}X_{+}^{a}|_{\tau
=0}
\end{equation}
where the coefficients $k^{\pm }$ and $l^{\pm }$ are functions of $\sigma
\equiv \frac{1}{2}(\sigma _{+}-\sigma _{-})$ to be determined, while
\begin{equation}
X_{+}^{a}|_{\tau =0}(\sigma )=\tfrac{1}{2}[1,\text{sgn}(\sigma )\,\mathbf{%
\hat{z}}]^{a}  \label{init1}
\end{equation}
and
\begin{equation}
X_{-}^{a}|_{\tau =0}(\sigma )=\tfrac{1}{2}[1,\sin (2\pi \sigma /L)\,\mathbf{%
\hat{x}}+\cos (2\pi \sigma /L)\,\mathbf{\hat{y}}]^{a}.  \label{init2}
\end{equation}

In the light-cone gauge, the constraint (\ref{const1}) reads simply
\begin{equation}
h_{ab}X_{+}^{a}X_{+}^{b}+2\eta _{ab}\delta X_{+}^{a}X_{+}^{b}=0\qquad \text{%
and}\qquad h_{ab}X_{-}^{a}X_{-}^{b}+2\eta _{ab}\delta X_{-}^{a}X_{-}^{b}=0
\label{lightconst}
\end{equation}
and (since $\eta _{ab}X_{-}^{a}X_{+}^{b}=\frac{1}{4}$) is satisfied on the
initial curve $\tau =0$ if
\begin{equation}
l^{+}=-2(h_{ab}X_{+}^{a}X_{+}^{b})|_{\tau =0}\qquad \text{and}\qquad
l^{-}=-2(h_{ab}X_{-}^{a}X_{-}^{b})|_{\tau =0}
\end{equation}
where $h_{ab}X_{+}^{a}X_{+}^{b}$ and $h_{ab}X_{-}^{a}X_{-}^{b}$ are known
functions of $u_{+}=(\tau +\sigma )/L$ given by (\ref{gpert1}) and (\ref
{gpert2}) respectively.

The initial configuration of the loop will be preserved if
\begin{equation}
0=\delta X_{\sigma }^{a}|_{\tau =0}=\delta X_{+}^{a}|_{\tau =0}-\delta
X_{-}^{a}|_{\tau =0}=(k^{+}-l^{-})X_{+}^{a}|_{\tau
=0}+(l^{+}-k^{-})X_{-}^{a}|_{\tau =0}
\end{equation}
at all points on the initial curve, and so $k^{+}=l^{-}$ and $l^{+}=k^{-}$.
Together with $\delta X^{a}|_{\tau =0}=0$, the initial data for the
perturbation $\delta X^{a}$ therefore have the form
\begin{equation}
\delta X_{+}^{a}|_{\tau =0}(\sigma )=\delta X_{-}^{a}|_{\tau =0}(\sigma
)=-2(h_{cd}X_{-}^{c}X_{-}^{d})|_{\tau =0}X_{+}^{a}|_{\tau
=0}-2(h_{cd}X_{+}^{c}X_{+}^{d})|_{\tau =0}X_{-}^{a}|_{\tau =0}.
\label{pertinit}
\end{equation}

It might seem at first glance that replacing $X_{\pm }^{a}|_{\tau =0}$ with $%
X_{\pm }^{a}|_{\tau =0}+\delta X_{\pm }^{a}|_{\tau =0}$ will induce a change
in the initial velocity of the loop. However, the effect of the changed
initial conditions is to simply realign the gauge coordinates in response to
the rotation of the light cones caused by the metric perturbation $h_{ab}$.
In the standard gauge, the tangent vector $X_{\tau }^{a}$ at any point $P$
on a given spacelike section $\tau ={}\,\!\!$constant of the world sheet is
uniquely determined by the tangent space at $P$ and the choice of gauge
coordinate $\sigma $ on the section, as $X_{\tau }^{a}$ is just the
future-pointing projection of the tangent space orthogonal to $X_{\sigma
}^{a}$, normalised so that $X_{\tau }^{2}=-X_{\sigma }^{2}$.

The equation of motion (\ref{eqmo3}) can now be integrated forward to give
\begin{equation}
\delta X_{+}^{a}(\sigma _{+},\sigma _{-})=\tfrac{1}{4}\int_{-\sigma
_{+}}^{\sigma _{-}}\alpha ^{a}(\sigma _{+},\theta )\,d\theta +\delta
X_{+}^{a}|_{\tau =0}(\sigma _{+})
\end{equation}
and
\begin{equation}
\delta X_{-}^{a}(\sigma _{+},\sigma _{-})=\tfrac{1}{4}\int_{-\sigma
_{-}}^{\sigma _{+}}\alpha ^{a}(\theta ,\sigma _{-})\,d\theta +\delta
X_{-}^{a}|_{\tau =0}(-\sigma _{-}).
\end{equation}
at all points with $\tau =\frac{1}{2}(\sigma _{+}+\sigma _{-})>0$.

Since
\begin{equation}
\int \alpha ^{t}du_{+}=8\mu L^{-1}\{\ln |W_{1}|-\func{Ci}(2\pi |W_{1}|)\},
\label{intt}
\end{equation}
\begin{eqnarray}
\int \alpha ^{L}du_{+} &=&8\mu L^{-1}[\{\func{Si}(4\pi u_{+})-\func{Si}(2\pi
D_{+})\}\cos 4\pi u_{+}  \nonumber \\
&&-\{\func{Ci}(4\pi u_{+})-\func{Ci}(2\pi D_{+})\}\sin 4\pi u_{+}-2\pi W_{1}]
\label{intL}
\end{eqnarray}
and
\begin{eqnarray}
&&\int \alpha ^{N}du_{+}=8\mu L^{-1}[\{\func{Si}(4\pi u_{+})-\func{Si}(2\pi
D_{+})\}\sin 4\pi u_{+}  \nonumber \\
&&+\{\func{Ci}(4\pi u_{+})-\func{Ci}(2\pi D_{+})\}\cos 4\pi u_{+}-\ln (4\pi
u_{+})+\ln D_{+}],  \label{intN}
\end{eqnarray}
it is possible to write down explicit expressions for $\delta
X_{+}^{a}$ and $\delta X_{-}^{a}$ at all points on the perturbed
trajectory. These in turn can be used to verify directly that the
gauge constraints (\ref {lightconst}) are satisfied at all ordinary
points on the trajectory.

For example,
\begin{eqnarray}
\tfrac{1}{4}\eta _{ab}X_{-}^{a}(\sigma _{-})\int_{-\sigma _{-}}^{\sigma
_{+}}\alpha ^{b}(\theta ,\sigma _{-})\,d\theta  &=&\tfrac{1}{8}\int_{-\sigma
_{-}}^{\sigma _{+}}\alpha ^{t}d\sigma _{+}-\tfrac{1}{8}\int_{-\sigma
_{-}}^{\sigma _{+}}\alpha ^{N}d\sigma _{+}  \nonumber \\
&=&-\tfrac{1}{2}(h_{cd}X_{-}^{c}X_{-}^{d})|_{\tau =0}(\sigma _{+})+\tfrac{1}{%
2}(h_{cd}X_{-}^{c}X_{-}^{d})|_{\tau =0}(-\sigma _{-})  \nonumber \\
&&
\end{eqnarray}
and
\begin{equation}
\eta _{ab}X_{-}^{a}(\sigma _{-})\delta X_{-}^{a}|_{\tau =0}(-\sigma _{-})=-%
\tfrac{1}{2}(h_{cd}X_{-}^{c}X_{-}^{d})|_{\tau =0}(-\sigma _{-})
\end{equation}
and so $\eta _{ab}X_{-}^{a}(\sigma _{-})\delta X_{-}^{a}(\sigma _{+},\sigma
_{-})=-\tfrac{1}{2}(h_{cd}X_{-}^{c}X_{-}^{d})(\sigma _{+})$ as required.
Similarly,
\begin{equation}
\tfrac{1}{4}\eta _{ab}X_{+}^{a}(\sigma _{+})\int_{-\sigma _{+}}^{\sigma
_{-}}\alpha ^{b}(\sigma _{+},\theta )\,d\theta =\tfrac{1}{4}\tau [\alpha
^{t}(\sigma _{+})-\text{sgn}(\sigma _{+})\alpha ^{z}(\sigma _{+})]=0
\end{equation}
and
\begin{equation}
\eta _{ab}X_{+}^{a}(\sigma _{+})\delta X_{+}^{a}|_{\tau =0}(\sigma _{+})=-%
\tfrac{1}{2}(h_{cd}X_{+}^{c}X_{+}^{d})|_{\tau =0}(\sigma _{+})
\end{equation}
and so $\eta _{ab}X_{+}^{a}(\sigma _{+})\delta X_{+}^{a}(\sigma _{+},\sigma
_{-})=-\tfrac{1}{2}(h_{cd}X_{+}^{c}X_{+}^{d})(\sigma _{+})$.

To track the cumulative change in the trajectory over a single oscillation
period $\Delta \tau =t_{p}\equiv L/2$, all that is necessary is to calculate
the position $X^{a}+\delta X^{a}$ of the loop and the tangent vectors $%
X_{\pm }^{a}+\delta X_{\pm }^{a}$ on the spacelike section $\tau =L/2$, turn
off the metric perturbation $h_{ab}$, then evolve the trajectory forwards in
flat space subject to the new initial conditions. The section $\tau =L/2$
corresponds to $\sigma _{\pm }=L/2\pm \sigma $, or equivalently (since $%
\alpha ^{a}$ and $\delta X_{\pm }^{a}$ are periodic functions of their
arguments with period $L$) to $\sigma _{\pm }=\pm (\sigma -L/2)$. So the
perturbed null tangent vectors after an oscillation period are
\begin{equation}
\delta X_{+}^{a}|_{\tau =L/2}(\sigma )=\tfrac{1}{4}\int_{-L/2}^{L/2}\alpha
^{a}(\sigma -L/2,\theta )\,d\theta +\delta X_{+}^{a}|_{\tau =0}(\sigma -L/2)
\end{equation}
and
\begin{equation}
\delta X_{-}^{a}|_{\tau =L/2}(\sigma )=\tfrac{1}{4}\int_{-L/2}^{L/2}\alpha
^{a}(\theta ,L/2-\sigma )\,d\theta +\delta X_{-}^{a}|_{\tau =0}(\sigma -L/2).
\end{equation}

The only dependence of $\alpha ^{a}$ on $\sigma _{-}$ is through the
explicit trigonometric functions in (\ref{accvec}), and so
\begin{equation}
\tfrac{1}{4}\int_{-L/2}^{L/2}\alpha ^{a}(\sigma -L/2,\theta )\,d\theta =%
\tfrac{1}{4}L[\alpha ^{t},\alpha ^{z}\mathbf{\hat{z}}]^{a},
\end{equation}
where $\alpha ^{t}$ and $\alpha ^{z}$ are now understood to be functions of $%
u_{+}=(\sigma -L/2)/L$. Also, in view of (\ref{intt}), (\ref{intL}) and (\ref
{intN}),
\begin{equation}
\tfrac{1}{4}\int_{-L/2}^{L/2}\alpha ^{t}\,d\sigma _{+}=\tfrac{1}{2}%
L\int_{0}^{1/2}\alpha ^{t}du_{+}=-4\mu [\gamma _{\text{E}}+\ln 2\pi -\func{Ci%
}(2\pi )]\approx -9.756\mu ,
\end{equation}
\begin{equation}
\tfrac{1}{4}\int_{-L/2}^{L/2}\alpha ^{L}d\sigma _{+}=-4\mu [2\pi -\func{Si}%
(2\pi )]\approx -19.460\mu
\end{equation}
and
\begin{equation}
\tfrac{1}{4}\int_{-L/2}^{L/2}\alpha ^{N}d\sigma _{+}=-4\mu [\gamma _{\text{E}%
}+\ln 2\pi -\func{Ci}(2\pi )]\approx -9.756\mu ,
\end{equation}
where $\gamma _{\text{E}}\approx 0.5772$ is Euler's constant as before. By
inspection, $\tfrac{1}{4}\int_{-L/2}^{L/2}\alpha ^{z}\,d\sigma _{+}=0$.

Hence, if $\Delta X_{\pm }^{a}(\sigma )=\delta X_{\pm }^{a}|_{\tau
=L/2}(\sigma )-\delta X_{\pm }^{a}|_{\tau =0}(\sigma -L/2)$ are the net
changes in the perturbations $\delta X_{\pm }^{a}$ over an oscillation
period then
\begin{equation}
\Delta X_{+}^{a}(\sigma )=\tfrac{1}{4}L[\alpha ^{t}(u_{+}),\alpha
^{z}(u_{+})\,\mathbf{\hat{z}}]^{a}
\end{equation}
and
\begin{equation}
\Delta X_{-}^{a}(\sigma )=[-\kappa ,(-\lambda \cos 2\pi u_{-}+\kappa \sin
2\pi u_{-})\mathbf{\hat{x}}-(\lambda \sin 2\pi u_{-}+\kappa \cos 2\pi u_{-})%
\mathbf{\hat{y}}]^{a},
\end{equation}
where $u_{+}=(\sigma -L/2)/L$ in the first equation and $u_{-}=-(\sigma
-L/2)/L$ in the second, while $\kappa =4\mu [\gamma +\ln 2\pi -\func{Ci}%
(2\pi )]$ and $\lambda =4\mu [2\pi -\func{Si}(2\pi )]$.

The perturbed solution can now be generated by integrating the Nambu-Goto
equation forward from the spacelike slice $\tau =L/2$ with the new initial
data $(X_{+}^{a}+\delta X_{+}^{a})|_{\tau =L/2}(\sigma )$ and $%
(X_{-}^{a}+\delta X_{-}^{a})|_{\tau =L/2}(\sigma )$, and $h_{ab}$ assumed to
be zero. Since $\delta X_{+}^{a}|_{\tau =0}=\delta X_{-}^{a}|_{\tau =0}$,
the tangent vector to the initial slice is:
\begin{eqnarray}
T^{a}(\sigma ) &=&(X_{\sigma }^{a}+\delta X_{\sigma }^{a})|_{\tau
=L/2}=(X_{+}^{a}+\delta X_{+}^{a})|_{\tau =L/2}-(X_{-}^{a}+\delta
X_{-}^{a})|_{\tau =L/2}  \nonumber \\
&=&\tfrac{1}{2}[0,\sin 2\pi u_{-}\,\mathbf{\hat{x}}-\cos 2\pi u_{-}\,\mathbf{%
\hat{y}}+\text{sgn}|u_{+}|\,\mathbf{\hat{z}}]^{a}+\Delta X_{+}^{a}-\Delta
X_{-}^{a}
\end{eqnarray}
The vector in the span of $(X_{+}^{a}+\delta X_{+}^{a})|_{\tau =L/2}$ and $%
(X_{-}^{a}+\delta X_{-}^{a})|_{\tau =L/2}$ that is orthogonal to $(X_{\sigma
}^{a}+\delta X_{\sigma }^{a})|_{\tau =L/2}$ (to linear order in $\delta
X_{\pm }^{a}$) is:
\begin{eqnarray}
N^{a}(\sigma ) &=&(X_{+}^{a}+\delta X_{+}^{a})|_{\tau
=L/2}+(X_{-}^{a}+\delta X_{-}^{a})|_{\tau =L/2}  \nonumber \\
&&-8(X_{-}\cdot \delta X_{-}\,X_{+}^{a})|_{\tau =L/2}-8(X_{+}\cdot \delta
X_{+}\,X_{-}^{a})|_{\tau =L/2}  \nonumber \\
&=&\tfrac{1}{2}[2,-\sin 2\pi u_{-}\,\mathbf{\hat{x}}+\cos 2\pi u_{-}\,%
\mathbf{\hat{y}}+\text{sgn}|u_{+}|\,\mathbf{\hat{z}}]^{a}  \nonumber \\
&&+\Delta X_{+}^{a}-8(X_{+}\cdot \Delta X_{+}\,X_{-}^{a})|_{\tau
=L/2}+\Delta X_{-}^{a}-8(X_{-}\cdot \Delta X_{-}\,X_{+}^{a})|_{\tau =L/2}
\nonumber \\
&&+\delta X_{+}^{a}|_{\tau =0}(Lu_{+})-8(X_{-}^{a}\,X_{+})|_{\tau =L/2}\cdot
\delta X_{+}|_{\tau =0}(Lu_{+})  \nonumber \\
&&+\delta X_{-}^{a}|_{\tau =0}(Lu_{+})-8(X_{+}^{a}\,X_{-})|_{\tau =L/2}\cdot
\delta X_{-}|_{\tau =0}(Lu_{+})
\end{eqnarray}
Note here from (\ref{pertinit}) that $X_{\pm }|_{\tau =L/2}\cdot \delta
X_{\pm }|_{\tau =0}(Lu_{+})=-\frac{1}{2}(h_{cd}X_{\pm }^{c}X_{\pm
}^{d})|_{\tau =0}(Lu_{+})$ and so the terms involving $\delta
X_{+}^{a}|_{\tau =0}$ and $\delta X_{-}^{a}|_{\tau =0}$ in $N^{a}$ all
cancel, leaving
\begin{equation}
N^{a}(\sigma )=\tfrac{1}{2}[2,-\sin 2\pi u_{-}\,\mathbf{\hat{x}}+\cos 2\pi
u_{-}\,\mathbf{\hat{y}}+\text{sgn}|u_{+}|\,\mathbf{\hat{z}}]^{a}+\Delta
X_{+}^{a}+\Delta X_{-}^{a}
\end{equation}
as $X_{+}\cdot \Delta X_{+}=X_{-}\cdot \Delta X_{-}=0$ on the slice $\tau
=L/2$.

Furthermore, the net displacement of the loop over an oscillation period can
be measured by tracking the motion of the centre-of-mass of the loop, which
for any constant-$\tau $ slice is defined to be
\begin{equation}
\mathbf{x}_{0}(\tau )=L^{-1}\int_{-L/2}^{L/2}[\mathbf{X}(\tau ,\sigma
)+\delta \mathbf{X}(\tau ,\sigma )]\text{\thinspace }d\sigma .
\end{equation}
In view of (\ref{posfunc2}), and the fact that $%
(h_{ab}X_{+}^{a}X_{+}^{b})|_{\tau =0}$ and $(h_{ab}X_{-}^{a}X_{-}^{b})|_{%
\tau =0}$ are periodic functions of $\sigma $ with period $L/2$ while $%
\mathbf{X}_{+}|_{\tau =0}$ and $\mathbf{X}_{-}|_{\tau =0}$ are anti-periodic
functions with the same period (and so $\mathbf{X}_{\tau }|_{\tau =0}$ and $%
\delta \mathbf{X}_{\tau }|_{\tau =0}$ are also anti-periodic),
\begin{equation}
\mathbf{x}_{0}(0)=\tfrac{1}{8}L\,\mathbf{\hat{z}}\qquad \text{and\qquad }%
\mathbf{x}_{0}^{\prime }(0)=\mathbf{0}.
\end{equation}

The subsequent motion of the centre-of-mass is governed by the equation
\begin{eqnarray}
&&\mathbf{x}_{0}^{\prime \prime }(\tau )=L^{-1}\int_{-L/2}^{L/2}(\mathbf{X}%
+\delta \mathbf{X}),_{\tau \tau }\text{\thinspace }d\sigma  \nonumber \\
&=&L^{-1}\int_{-L/2}^{L/2}[(\alpha ^{L}\cos 2\pi u_{-}-\alpha ^{N}\sin 2\pi
u_{-})\mathbf{\hat{x}}+(\alpha ^{L}\sin 2\pi u_{-}+\alpha ^{N}\cos 2\pi
u_{-})\mathbf{\hat{y}}+\alpha ^{z}\mathbf{\hat{z}}]\text{\thinspace }d\sigma
\nonumber \\
&&
\end{eqnarray}
where the second line follows because $(X^{a}+\delta X^{a}),_{\tau \tau
}=\alpha ^{a}+(X^{a}+\delta X^{a}),_{\sigma \sigma }$ and $\mathbf{X}%
_{\sigma }+\delta \mathbf{X}_{\sigma }$ is a periodic function of $\sigma $
on $[-L/2,L/2]$. For fixed values of $\tau $ the acceleration components $%
\alpha ^{L}$ and $\alpha ^{N}$ are periodic functions of $\sigma $ with
period $L/2$, whereas $\alpha ^{z}$, $\cos 2\pi u_{-}$ and $\sin 2\pi u_{-}$
are anti-periodic with period $L/2$. So $\mathbf{x}_{0}^{\prime \prime
}(\tau )=\mathbf{0}$, and the centre-of-mass remains unperturbed by the
back-reaction, a feature which is in any case obvious from the symmetry of
the ACO loop.

Let $X^{a}(\tau ,\sigma )$ now denote the flat-space solution generated by
the initial conditions $X_{\sigma }^{a}=T^{a}(\sigma )$ and $X_{\tau
}^{a}=N^{a}(\sigma )$ on the spacelike slice $\tau =L/2$. The equation of
motion $X^{a},_{\tau \tau }=X^{a},_{\sigma \sigma }$ is then satisfied by
taking
\begin{equation}
X_{+}^{a}=\tfrac{1}{2}[N^{a}(\sigma _{+})+T^{a}(\sigma _{+})]
\end{equation}
and
\begin{equation}
X_{-}^{a}=\tfrac{1}{2}[N^{a}(-\sigma _{-})-T^{a}(-\sigma _{-})]
\end{equation}
with $\sigma _{\pm }=\tau \pm \sigma $ as usual. The position vector $%
\mathbf{X}$ itself can be found on each constant-$\tau $ section by
integrating the equation $\mathbf{X}_{\sigma }=\mathbf{X}_{+}-\mathbf{X}_{-}$
subject to the constraint $\mathbf{x}_{0}(\tau )=\tfrac{1}{8}L\,\mathbf{\hat{%
z}}$.

However, there is no simple relationship between the gauge coordinate $\tau $
in the new solution and the Minkowski time coordinate $t=X^{0}$, as
\begin{equation}
X_{\tau }^{0}|_{\tau =L/2}=N^{0}(\sigma )=1+\tfrac{1}{4}L\alpha
^{t}(u_{+})-\kappa \neq 1.
\end{equation}
In order to facilitate comparison between successive stages in the
evaporation of the loop, it is useful to realign the coordinates $\sigma
_{+} $ and $\sigma _{-}$ by replacing them with two new coordinates
\begin{equation}
\hat{\sigma}_{+}=\int_{0}^{\sigma _{+}}[N^{0}(\sigma )+T^{0}(\sigma
)]\,d\sigma =\sigma _{+}+\tfrac{1}{2}L\int_{0}^{\sigma _{+}}\alpha
^{t}(\theta /L)\,d\theta
\end{equation}
and
\begin{equation}
\hat{\sigma}_{-}=-\int_{-\sigma _{-}}^{0}[N^{0}(\sigma
)-T^{0}(\sigma )]\,d\sigma =(1-2\kappa )\sigma _{-},
\end{equation}
defined so that $\partial X^{t}/\partial \hat{\sigma}_{+}=\partial
X^{t}/\partial \hat{\sigma}_{+}=\frac{1}{2}$ at all points on the
new trajectory. Note that because $\hat{\sigma}_{\pm }$ is a
function of $\sigma _{\pm }$ only, the form of the flat-space
equation of motion $X^{a},_{+-}=0$ is preserved.

Furthermore, since $X_{\pm }^{a}$ and $\Delta X_{\pm }^{a}$ are periodic
functions of $\sigma _{\pm }$ with period $L$, the net change in $\hat{\sigma%
}_{+}$ or $\hat{\sigma}_{-}$ over a parametric period is:
\begin{equation}
\Delta \hat{\sigma}_{\pm }=L+\tfrac{1}{2}\int_{0}^{L}\int_{0}^{L}\alpha
^{t}d\sigma _{+}\,d\sigma _{-}=L-2\kappa L.
\end{equation}
Thus, in a gauge coordinate system aligned with the Minkowski time
coordinate $t$, the parametric period changes over a single oscillation
period $t_{p}$ from $L$ to $L^{\prime }=L+\Delta L$, where
\begin{equation}
\Delta L=-2\kappa L\approx -19.501\mu L.
\end{equation}
Since the total energy of the string loop is $E=\mu L$, the energy radiated
by the loop over a complete parametric period $2t_{p}$ is $\Delta E=2\mu
\Delta L\approx -39.002\mu ^{2}L$, in agreement with the original
calculation of Allen, Casper and Ottewill \cite{ACO1}.

In terms of the new gauge coordinates $\hat{\sigma}_{+}$ and $\hat{\sigma}%
_{-}$, the initial data on the constant-time slice $\tau =L/2$ becomes
\begin{eqnarray}
&&\partial X^{a}/\partial \hat{\sigma}_{+}=(X_{+}^{a}\tfrac{d\sigma _{+}}{d%
\hat{\sigma}_{+}})|_{\tau =L/2}  \nonumber \\
&=&\{\tfrac{1}{2}[1,\text{sgn}(\sigma -L/2)\,\mathbf{\hat{z}}]^{a}+\tfrac{1}{%
4}L[\alpha ^{t}(u_{+}),\alpha ^{z}(u_{+})\,\mathbf{\hat{z}}]^{a}\}\{1+\tfrac{%
1}{2}L\alpha ^{t}(u_{+})\}^{-1}  \nonumber \\
&=&\tfrac{1}{2}[1,\text{sgn}(\sigma -L/2)\,\mathbf{\hat{z}}]^{a}
\label{new1}
\end{eqnarray}
as $\alpha ^{z}(u_{+})={}\,\!\!$sgn$(u_{+})\alpha ^{t}(u_{+})$; and
\begin{eqnarray}
\partial X^{a}/\partial \hat{\sigma}_{-} &=&\tfrac{1}{2}[1,\{-2\lambda
(1-2\kappa )^{-1}\cos 2\pi u_{-}-\sin 2\pi u_{-}\}\mathbf{\hat{x}}  \nonumber
\\
&&+\{-2\lambda (1-2\kappa )^{-1}\sin 2\pi u_{-}+\cos 2\pi u_{-})\}\mathbf{%
\hat{y}}]^{a}.  \label{new2}
\end{eqnarray}
The final step in demonstrating that the ACO loop evaporates by self-similar
shrinkage is to show that the tangent vectors (\ref{new1}) and (\ref{new2})
are the same functions (to order $\mu $) of $\hat{\sigma}_{+}$ and $\hat{%
\sigma}_{-}$, respectively, as $X_{+}^{a}$ and $X_{-}^{a}$ are of $\sigma $
in the initial configuration (\ref{init1}) and (\ref{init2}), save for a
change in parametric period from $L$ to $L^{\prime }$, and a rotational
phase shift in (\ref{new2}).

In the case of (\ref{new1}) note that sgn$(\sigma -L/2)$ is the restriction
to the interval $(0,L)$ of its $L$-periodic extension. As $\sigma $ varies
from $0$ to $L$ on the constant-time slice $\tau =L/2$, $\sigma _{+}$ varies
from $L/2$ to $3L/2$ and $\hat{\sigma}_{+}$ varies from $L/2-\kappa L$ to $%
3L/2-3\kappa L$. Thus, the periodic extension of
\begin{equation}
\text{sgn}(\sigma -L/2)\equiv \text{sgn}(\sigma _{+}-L)
\end{equation}
with period $L$ is the same as the periodic extension of
\begin{equation}
\text{sgn}(\hat{\sigma}_{+}-L+2\kappa L)\equiv \text{sgn}(\hat{\sigma}%
_{+}-L^{\prime })
\end{equation}
on $(L/2-\kappa L,3L/2-3\kappa L)$ with period $L^{\prime }=(1-2\kappa )L$.
But this in turn is the same as the $L^{\prime }$-periodic extension of sgn$(%
\hat{\sigma}_{+})$ on $(-L^{\prime }/2,L^{\prime }/2)$, and
\begin{equation}
\partial X^{a}/\partial \hat{\sigma}_{+}=\tfrac{1}{2}[1,\text{sgn}(\hat{%
\sigma}_{+})\,\mathbf{\hat{z}}]^{a},
\end{equation}
as required.

Referring now to (\ref{new2}), the constants $\kappa $ and $\lambda $ are
both of order $\mu $, while
\begin{equation}
u_{-}=\sigma _{-}/L=\hat{\sigma}_{-}L^{-1}(1-2\kappa )^{-1}=\hat{\sigma}%
_{-}/L^{\prime }
\end{equation}
on the slice $\tau =L/2$. So
\begin{equation}
\partial X^{a}/\partial \hat{\sigma}_{-}\approx \tfrac{1}{2}[1,-(2\lambda
\cos 2\pi \hat{u}_{-}+\sin 2\pi \hat{u}_{-})\mathbf{\hat{x}}+(-2\lambda \sin
2\pi \hat{u}_{-}+\cos 2\pi \hat{u}_{-})\mathbf{\hat{y}}]^{a}
\end{equation}
where $\hat{u}_{-}=\hat{\sigma}_{-}/L^{\prime }$. Note here that
\begin{equation}
\eta _{ab}\partial X^{a}/\partial \hat{\sigma}_{-}\partial X^{b}/\partial
\hat{\sigma}_{-}=-\lambda ^{2}
\end{equation}
and therefore $\partial X^{a}/\partial \hat{\sigma}_{-}$ remains a null
vector to first order in $\mu $. In fact,
\begin{equation}
-(2\lambda \cos 2\pi \hat{u}_{-}+\sin 2\pi \hat{u}_{-})=-A\sin (2\pi \hat{u}%
_{-}+\psi )
\end{equation}
and
\begin{equation}
-2\lambda \sin 2\pi \hat{u}_{-}+\cos 2\pi \hat{u}_{-}=A\cos (2\pi \hat{u}%
_{-}+\psi )
\end{equation}
where
\begin{equation}
\psi =\tan ^{-1}(2\lambda )\approx 2\lambda \text{\qquad and\qquad }%
A=(1+4\lambda ^{2})^{1/2}\approx 1
\end{equation}
to first order in $\mu $. It follows that
\begin{equation}
\partial X^{a}/\partial \hat{\sigma}_{-}\approx \tfrac{1}{2}[1,-\sin (2\pi
\hat{u}_{-}+2\lambda )\mathbf{\hat{x}}+\cos (2\pi \hat{u}_{-}+2\lambda )%
\mathbf{\hat{y}}]^{a}
\end{equation}
and $\partial X^{a}/\partial \hat{\sigma}_{-}$ is the same function of $\hat{%
u}_{-}$ as $X^{a},_{-}$ is of $\sigma /L$ in (\ref{init2}), apart from the
phase shift of $2\lambda \approx 38.92\mu $.

Note finally that if (\ref{new1}) and (\ref{new2}) are used as initial data
for a new oscillation of the loop, with $\hat{\tau}$ rezeroed so that the
initial surface $t=L/2$ corresponds to $\hat{\tau}=0$, then
\begin{eqnarray}
&&\partial X^{a}/\partial \hat{\sigma}|_{\hat{\tau}=0}=(\partial
X^{a}/\partial \hat{\sigma}_{+}-\partial X^{a}/\partial \hat{\sigma}_{-})|_{%
\hat{\tau}=0}  \nonumber \\
&=&\tfrac{1}{2}[1,\text{sgn}(\hat{\sigma})\,\mathbf{\hat{z}}]^{a}-\tfrac{1}{2%
}[1,-\sin (-2\pi \hat{\sigma}/L^{\prime }+2\lambda )\mathbf{\hat{x}}+\cos
(-2\pi \hat{\sigma}/L^{\prime }+2\lambda )\mathbf{\hat{y}}]^{a}  \nonumber \\
&&
\end{eqnarray}
and so
\begin{equation}
X^{a}(\hat{\sigma})|_{\hat{\tau}=0}=[0,(\tfrac{1}{2}|\hat{\sigma}|+\tfrac{1}{%
8}|\Delta L|)\,\mathbf{\hat{z}}]^{a}+\tfrac{L^{\prime }}{4\pi }[0,\cos
(-2\pi \hat{\sigma}/L^{\prime }+2\lambda )\mathbf{\hat{x}}+\sin (-2\pi \hat{%
\sigma}/L^{\prime }+2\lambda )\mathbf{\hat{y}}]^{a}
\end{equation}
Hence, the kink point at $\hat{\sigma}=0$ corresponds to the spacetime point
\begin{equation}
X^{a}=[L/2,\tfrac{1}{8}|\Delta L|\,\mathbf{\hat{z}}]^{a}+\tfrac{L^{\prime }}{%
4\pi }[0,\cos (2\lambda )\mathbf{\hat{x}}+\sin (2\lambda )\mathbf{\hat{y}}%
]^{a}
\end{equation}
which is in the same spatial position as the kink point was in the
unperturbed trajectory at the slightly later time $\tau =-\sigma =\lambda
L/2\pi $ (except for a uniform contraction towards the centre-of-mass point).

This observation throws some light on a question that was raised in Section
6.11 of \cite{Anderson2}, namely whether gravitational radiation in the
weak-field approximation would be expected to force the overall pattern of a
loop (and in particular salient features such as kinks and cusps) to advance
or precess. In the case of the ACO\ loop it can be seen that the pattern
advances. The phase shift here is due entirely to the form of the lateral
acceleration $\alpha ^{L}$. If the lateral acceleration is everywhere
inwards, and the $x$-$y$ position vector is a single-mode harmonic function
of $\sigma _{-}$ (or $\sigma _{+}$), then it is easily seen that advance of
the pattern will always occur. But advance of the pattern does not seem to
be a general feature of all loops.

\section{Conclusion}

In this paper I have generated explicit formulae for the weak-field metric
perturbations induced by an Allen-Casper-Ottewill (ACO) loop of cosmic
string and calculated the corresponding weak-field back-reaction
acceleration vector. Although the acceleration vector diverges at the two
kink points on the loop, it turns out that the net acceleration and
radiative energy loss of the loop are finite. Using a method first described
by Quashnock and Spergel, I have determined the net effect of the
back-reaction on the motion of the loop over a single oscillation period,
and shown that the new, perturbed trajectory of the loop is identical to the
original, save for a uniform contraction of scale and a small rotational
phase shift. The ACO loop, which is rigidly rotating and has the lowest
radiative efficiency of any known loop solution, therefore evolves by
self-similar evaporation at the weak-field level. This in turn means that,
in any ensemble of loops with a given energy $E$, the ACO\ loops (and
presumably any loops similar in shape) will be the most stable and
longest-lived.

The conclusions drawn here depend of course on the assumption that the
Quashnock-Spergel method and the weak-field approximation accurately model
the secular evolution of the ACO\ loop. Although there is no reason for
suspecting that either approximation introduces spurious features into the
calculation, it would be a useful check to be able to generate a \emph{%
continuously} self-similar solution to the back-reaction equations at the
weak-field level or (even better) within the full framework of general
relativity.

I would like to thank the referee for some helpful suggestions.

\appendix

\section{Calculation of the self-acceleration vector}

As was mentioned in Section \ref{pertsec}, if the field point $[\bar{t},%
\mathbf{\bar{x}}]$ is a point $[\bar{\tau},\mathbf{\bar{X}}]$ on the string
the contour $\Gamma $ in the integral (\ref{weakh3}) for $h_{ab}$ divides
naturally into three segments, which have been labelled $\Gamma ^{-}$, $%
\Gamma ^{0}$ and $\Gamma ^{+}$. When $u_{+}>0$ the contour $\Gamma $
consists first of $\Gamma ^{-}$ then $\Gamma ^{0}$ then $\Gamma ^{+}$, with $%
(v_{+},v_{-})$ ranging from $(-\tfrac{1}{2},u_{-}+W_{1}+1)$ to $(0,u_{-})$
to $(u_{+},u_{-})$ to $(\tfrac{1}{2},u_{-}+W_{1})$ along the segments. On
the other hand, if $u_{+}<0$ then the order of the segments is first $\Gamma
^{0}$ then $\Gamma ^{-}$ then $\Gamma ^{+}$, with $(v_{+},v_{-})$ ranging
from $(-\tfrac{1}{2},u_{-})$ to $(u_{+},u_{-})$ to $(0,u_{-}+W_{0})$ to $(%
\tfrac{1}{2},u_{-}-1)$. The divergence in $h_{ab}$ at an ordinary point on
the string is due to the contribution near $(v_{+},v_{-})=(u_{+},u_{-})$,
which lies on the boundary between $\Gamma ^{0}$ and $\Gamma ^{+}$ if $%
u_{+}>0$, and on the boundary between $\Gamma ^{0}$ and $\Gamma ^{-}$ if $%
u_{+}<0$.

Now, if the field point $[\bar{t},\mathbf{\bar{x}}]$ lies off the world
sheet $\mathbf{T}$, the line integral for $h_{ab}$ takes the form previously
given in (\ref{metpert}), namely
\begin{equation}
h_{ab}(\bar{t},\mathbf{\bar{x}})=-4\mu L\int_{V_{1}}^{V_{0}}(\Delta
_{+}^{-1}\Psi _{ab})|_{s=1}\,dv_{-}-4\mu L\int_{V_{0}}^{V_{1}+1}(\Delta
_{+}^{-1}\Psi _{ab})|_{s=-1}\,dv_{-}  \label{metpert2}
\end{equation}
where the limit functions $V_{0}$ and $V_{1}$ depend on $\bar{x}^{c}=[\bar{t}%
,\mathbf{\bar{x}}]^{c}$ through equation (\ref{roots}), while the integrand $%
\Delta _{+}^{-1}\Psi _{ab}$ depends on $\bar{x}^{c}$, $v_{-}$ and (in
principle) $v_{+}$, which is itself a function of $\bar{x}^{c}$ and $v_{-}$
on $\Gamma $ through (\ref{gamma}). However, the dependence of $\Psi _{ab}$
and $\Delta _{+}$ on $v_{+}$ involves a dependence on $s=\,$sgn$(v_{+})$
only, and $s$ is constant on the two sub-integrals in (\ref{metpert2}).

The spacetime derivatives of the metric perturbations therefore have the
form
\begin{eqnarray}
&&h_{ab},_{c}(\bar{t},\mathbf{\bar{x}})=-4\mu
L\{H_{abc}(0^{+},V_{0})-H_{abc}(\tfrac{1}{2},V_{1})+H_{abc}(-\tfrac{1}{2}%
,V_{1}+1)-H_{abc}(0^{-},V_{0})\}  \nonumber \\
&&+4\mu L\int_{V_{1}}^{V_{0}}(\Delta _{+}^{-2}\Psi _{ab}\Delta
_{+},_{c})|_{s=1}\,dv_{-}+4\mu L\int_{V_{0}}^{V_{1}+1}(\Delta _{+}^{-2}\Psi
_{ab}\Delta _{+},_{c})|_{s=-1}\,dv_{-}  \label{hderiv}
\end{eqnarray}
where
\begin{equation}
\Delta _{+},_{c}=[1,-s\mathbf{\hat{z}}]_{c}
\end{equation}
and
\begin{equation}
H_{abc}(v_{+},V)=(\Delta _{+}^{-1}\Psi _{ab}V,_{c})|_{v_{-}=V}.
\end{equation}
Furthermore, differentiation of the light-cone condition $\frac{1}{2}%
L(v_{+}+V)=\bar{t}-|\mathbf{\bar{x}}-\mathbf{X}(Lv_{+},LV)|$ with $v_{+}$
constant gives
\begin{equation}
V,_{c}=2L^{-1}\Delta _{-}^{-1}[\bar{t}-\tau ,-(\mathbf{\bar{x}}-\mathbf{X}%
)]_{c}  \label{Vderiv}
\end{equation}
with $\tau =\frac{1}{2}L(v_{+}+v_{-})$ and $\Delta _{-}=\bar{t}-\tau -2(%
\mathbf{\bar{x}}-\mathbf{X})\cdot \mathbf{X}_{-}$ as before.

In view of (\ref{accel2}), (\ref{hderiv}) and (\ref{Vderiv}), the
acceleration vector $\alpha ^{a}$ can be written in the form
\begin{eqnarray}
\alpha ^{a}(u_{+},u_{-}) &=&-8\mu L\{A^{a}(0^{+},V_{0})-A^{a}(\tfrac{1}{2}%
,V_{1})+A^{a}(-\tfrac{1}{2},V_{1}+1)-A^{a}(0^{-},V_{0})\}  \nonumber \\
&&+8\mu L\int_{V_{1}}^{V_{0}}B^{a}|_{s=1}\,dv_{-}+8\mu
L\int_{V_{0}}^{V_{1}+1}B^{a}|_{s=-1}\,dv_{-},  \label{accel3}
\end{eqnarray}
where
\begin{equation}
A^{a}(v_{+},V)=-\Delta _{+}^{-1}\{(\Psi _{b}^{a}\bar{X}_{+}^{b}\,\bar{X}%
_{-}^{c}+\Psi _{b}^{a}\bar{X}_{-}^{b}\,\bar{X}_{+}^{c})V,_{c}-\Psi _{bc}\bar{%
X}_{+}^{b}\,\bar{X}_{-}^{c}\,\eta ^{ad}V,_{d}\}|_{v_{-}=V}  \label{Aa}
\end{equation}
and
\begin{equation}
B^{a}=-\Delta _{+}^{-2}\{(\Psi _{b}^{a}\bar{X}_{+}^{b}\,\bar{X}_{-}^{c}+\Psi
_{b}^{a}\bar{X}_{-}^{b}\,\bar{X}_{+}^{c})\Delta _{+},_{c}-\Psi _{bc}\bar{X}%
_{+}^{b}\,\bar{X}_{-}^{c}\,\eta ^{ad}\Delta _{+},_{d}\}  \label{Ba}
\end{equation}
and again it is understood that the limit $[\bar{t},\mathbf{\bar{x}}%
]\rightarrow [\bar{\tau},\mathbf{\bar{X}}]$ is taken on the right-hand side
of (\ref{accel3}).

Now,
\begin{equation}
\bar{X}_{+}^{b}=\tfrac{1}{2}[1,\bar{s}\mathbf{\hat{z}}]^{b}\qquad \text{%
and\qquad }\bar{X}_{-}^{c}=\tfrac{1}{2}[1,-\sin (2\pi u_{-}\,)\mathbf{\hat{x}%
}+\cos (2\pi u_{-})\mathbf{\hat{y}}]^{c},  \label{derivs}
\end{equation}
while the components of $\Psi _{ab}$ are given in (\ref{source2}). Hence,
\begin{equation}
\Psi _{b}^{a}\bar{X}_{+}^{b}=\tfrac{1}{4}(1-s\bar{s})[1,-\sin (2\pi v_{-})%
\mathbf{\hat{x}}+\cos (2\pi v_{-})\mathbf{\hat{y}}+s\mathbf{\hat{z}}]^{a}
\label{term1}
\end{equation}
\begin{eqnarray}
&&\Psi _{b}^{a}\bar{X}_{-}^{b}=\tfrac{1}{4}\{1-\cos 2\pi (v_{-}-u_{-})\}[1,s%
\mathbf{\hat{z}}]^{a}  \nonumber \\
&&+\tfrac{1}{4}[0,-\{\sin (2\pi v_{-})-\sin (2\pi u_{-})\}\mathbf{\hat{x}}%
+\{\cos (2\pi v_{-})-\cos (2\pi u_{-})\}\mathbf{\hat{y}}]^{a}  \label{term2}
\end{eqnarray}
and
\begin{equation}
\Psi _{bc}\bar{X}_{+}^{b}\,\bar{X}_{-}^{c}=\tfrac{1}{8}(1-s\bar{s})\{1-\cos
2\pi (v_{-}-u_{-})\}.  \label{term3}
\end{equation}

Furthermore,
\begin{equation}
\bar{X}_{-}^{c}\Delta _{+},_{c}=\tfrac{1}{2}\qquad \text{and\qquad }\bar{X}%
_{+}^{c}\Delta _{+},_{c}=\tfrac{1}{2}(1-s\bar{s}),
\end{equation}
while the general limiting values of $\Delta _{+}$ and $\Delta _{-}$ at $[%
\bar{t},\mathbf{\bar{x}}]=[\bar{\tau},\mathbf{\bar{X}}]$ are
\begin{equation}
\Delta _{+}(v_{+},v_{-})=\tfrac{1}{2}L\{u_{+}(1-s\bar{s})+u_{-}-v_{-}\}
\label{delplus}
\end{equation}
and
\begin{equation}
\Delta _{-}(v_{+},v_{-})=\tfrac{1}{2}L\{u_{+}-v_{+}+u_{-}-v_{-}+\tfrac{1}{%
2\pi }\sin 2\pi (v_{-}-u_{-})\}.  \label{delmin}
\end{equation}

For the moment, attention will be restricted to the case where $u_{+}>0$ and
so $\bar{s}=1$. It was seen in Section \ref{pertsec} that the integral for $%
h_{ab}(\bar{t},\mathbf{\bar{x}})$ diverges in the limit as $[\bar{t},\mathbf{%
\bar{x}}]\rightarrow [\bar{\tau},\mathbf{\bar{X}}]$, because $%
V_{0}\rightarrow u_{-}$ and so $\Delta _{+}(0^{+},V_{0})\rightarrow 0$, as
is evident from (\ref{delplus}). It is clear from (\ref{Aa}) and (\ref{Ba})
that there may be similar divergences in $A^{a}(0^{+},V_{0})$ and $%
\int_{V_{1}}^{V_{0}}B^{a}|_{s=1}\,dv_{-}$. However, it turns out that both
functions are in fact convergent, and therefore that $\alpha ^{a}$ is
well-defined.

To see this, note first that if $u_{+}>0$ and $v_{+}=0^{+}$ then $s=\bar{s}%
=1 $ and so the terms (\ref{term1}) and (\ref{term3}) vanish identically.
Therefore
\begin{equation}
A^{a}(0^{+},V_{0})=-(\Delta _{+}^{-1}\Psi _{b}^{a}\bar{X}_{-}^{b}\,\bar{X}%
_{+}^{c}V,_{c})|_{(0^{+},V_{0})}
\end{equation}
where, at any field point $[\bar{t},\mathbf{\bar{x}}]$,
\begin{eqnarray}
\,\bar{X}_{+}^{c}V,_{c}|_{(0^{+},V_{0})} &=&L^{-1}\Delta _{-}^{-1}\{\bar{t}-%
\tfrac{1}{2}L(v_{+}+v_{-})-\bar{s}(\bar{z}-\tfrac{1}{2}Lsv_{+})%
\}|_{(0^{+},V_{0})}  \nonumber \\
&=&L^{-1}(\Delta _{-}^{-1}\Delta _{+})|_{(0^{+},V_{0})}
\end{eqnarray}
and so
\begin{equation}
A^{a}(0^{+},V_{0})=-L^{-1}(\Delta _{-}^{-1}\Psi _{b}^{a}\bar{X}%
_{-}^{b})|_{(0^{+},V_{0})}
\end{equation}
In the limit as $[\bar{t},\mathbf{\bar{x}}]\rightarrow [\bar{\tau},\mathbf{%
\bar{X}}]$, it can be seen from (\ref{delmin}) that $(\Delta
_{-}^{-1})|_{(0^{+},V_{0})}\rightarrow 2(Lu_{+})^{-1}$ and from (\ref{term2}%
) that $(\Psi _{b}^{a}\bar{X}_{-}^{b})|_{(0^{+},V_{0})}$ vanishes. Hence, $%
A^{a}(0^{+},V_{0})$ also vanishes in the limit.

Similarly, because $\Psi _{b}^{a}\bar{X}_{+}^{b}$ and $\bar{X}_{+}^{c}\Delta
_{+},_{c}$ and $\Psi _{cd}\bar{X}_{+}^{b}\bar{X}_{-}^{c}$ are all
identically zero when $s=\bar{s}$, it follows that $B^{a}|_{s=1}$ is also
identically zero, and therefore that $\int_{V_{1}}^{V_{0}}B^{a}|_{s=1}%
\,dv_{-}=0$ at any field point $[\bar{t},\mathbf{\bar{x}}]$. Furthermore,
all other limiting values of $\Delta _{+}$ and $\Delta _{-}$ are non-zero,
as $V_{1}\rightarrow W_{1}(u_{+})+u_{-}$ in the limit as $[\bar{t},\mathbf{%
\bar{x}}]\rightarrow [\bar{\tau},\mathbf{\bar{X}}]$ and so
\[
\Delta _{+}\rightarrow \tfrac{1}{2}L(2u_{+}-1-W_{1})\text{\qquad and\qquad }%
\Delta _{-}\rightarrow \tfrac{1}{2}L(u_{+}-\tfrac{1}{2}-W_{1}+\tfrac{1}{2\pi
}\sin 2\pi W_{1})
\]
at $(-\tfrac{1}{2},V_{1}+1)$;
\[
\Delta _{+}\rightarrow Lu_{+}\text{\qquad and\qquad }\Delta _{-}\rightarrow
\tfrac{1}{2}Lu_{+}
\]
at $(0^{-},V_{0})$; and
\[
\Delta _{+}\rightarrow -\tfrac{1}{2}LW_{1}\text{\qquad and\qquad }\Delta
_{-}\rightarrow \tfrac{1}{2}L(u_{+}-\tfrac{1}{2}-W_{1}+\tfrac{1}{2\pi }\sin
2\pi W_{1})
\]
at $(\tfrac{1}{2},V_{1})$. It is easily verified that these limiting values
are all positive if $u_{+}\in (0,\frac{1}{2})$.

The limiting values of the three non-zero $M_{b}^{a}A^{b}$ terms
contributing to the co-rotating acceleration vector $M_{b}^{a}\alpha ^{b}$
through (\ref{accel3}) can now be found by direct substitution. If $u_{+}>0$
then
\begin{equation}
M_{b}^{a}A^{b}(\tfrac{1}{2},V_{1})\rightarrow -\tfrac{1}{2}%
L^{-2}D_{-}^{-1}\{(1-\cos 2\pi W_{1})[1,\mathbf{\hat{z}}]^{a}+[0,-\sin 2\pi
W_{1}\mathbf{\hat{x}}-(1-\cos 2\pi W_{1})\mathbf{\hat{y}}]^{a}\}  \label{A1}
\end{equation}
\begin{equation}
M_{b}^{a}A^{b}(0^{-},V_{0})\rightarrow -\tfrac{1}{2}L^{-2}u_{+}^{-1}[1,%
\mathbf{\hat{y}}-\mathbf{\hat{z}}]^{a}  \label{A2}
\end{equation}
and
\begin{eqnarray}
&&M_{b}^{a}A^{b}(-\tfrac{1}{2},V_{1}+1)\rightarrow \tfrac{1}{2}%
L^{-2}D_{-}^{-1}(1-\cos 2\pi W_{1})[1,\mathbf{\hat{z}}]^{a}  \nonumber \\
&&-L^{-2}D_{+}^{-1}[1,-\sin 2\pi W_{1}\mathbf{\hat{x}}+\cos 2\pi W_{1}%
\mathbf{\hat{y}}-\mathbf{\hat{z}}]^{a}  \nonumber \\
&&-\tfrac{1}{2\pi }L^{-2}(1-\cos 2\pi W_{1})D_{+}^{-1}D_{-}^{-1}[0,-(1-\cos
2\pi W_{1})\mathbf{\hat{x}}+\sin 2\pi W_{1}\mathbf{\hat{y}}]^{a}  \nonumber
\\
&&+\tfrac{1}{2}L^{-2}D_{+}^{-1}D_{-}^{-1}W_{1}[0,-\sin 2\pi W_{1}\mathbf{%
\hat{x}}-(1-\cos 2\pi W_{1})\mathbf{\hat{y}}]^{a}  \label{A3}
\end{eqnarray}
where
\begin{equation}
D_{+}=2u_{+}-1-W_{1}\text{\qquad and\qquad }D_{-}=u_{+}-\tfrac{1}{2}-W_{1}+%
\tfrac{1}{2\pi }\sin 2\pi W_{1}
\end{equation}

The one remaining contribution to $M_{b}^{a}\alpha ^{b}$ is $%
\int_{V_{0}}^{V_{1}+1}M_{b}^{a}B^{b}|_{s=-1}\,dv_{-}$. If $s=-1$ then
\begin{equation}
\Psi _{b}^{a}\bar{X}_{+}^{b}\,\bar{X}_{-}^{c}\Delta _{+},_{c}=\tfrac{1}{4}[%
1,-\sin (2\pi v_{-})\mathbf{\hat{x}}+\cos (2\pi v_{-})\mathbf{\hat{y}}-%
\mathbf{\hat{z}}]^{a}
\end{equation}
\begin{eqnarray}
&&\Psi _{b}^{a}\bar{X}_{-}^{b}\,\bar{X}_{+}^{c}\Delta _{+},_{c}=\tfrac{1}{4}%
\{1-\cos 2\pi (v_{-}-u_{-})\}[1,-\mathbf{\hat{z}}]^{a}  \nonumber \\
&&+\tfrac{1}{4}[0,-\{\sin (2\pi v_{-})-\sin (2\pi u_{-})\}\mathbf{\hat{x}}%
+\{\cos (2\pi v_{-})-\cos (2\pi u_{-})\}\mathbf{\hat{y}}]^{a}
\end{eqnarray}
and
\begin{equation}
\Psi _{bc}\bar{X}_{+}^{b}\bar{X}_{-}^{c}\eta ^{ad}\Delta _{+},_{d}=\tfrac{1}{%
4}\{1-\cos 2\pi (v_{-}-u_{-})\}[1,-\mathbf{\hat{z}}]^{a},
\end{equation}
and so
\begin{eqnarray}
&&M_{b}^{a}B^{b}|_{s=-1}=-L^{-2}(v_{-}-2\bar{t}/L-2\bar{z}/L)^{-2}[1,-%
\mathbf{\hat{y}}-\mathbf{\hat{z}}]^{a}  \nonumber \\
&&+2L^{-2}(v_{-}-2\bar{t}/L-2\bar{z}/L)^{-2}[0,\sin 2\pi (v_{-}-u_{-})%
\mathbf{\hat{x}}-\cos 2\pi (v_{-}-u_{-})\mathbf{\hat{y}]}^{a}  \nonumber \\
&&
\end{eqnarray}
Hence,
\begin{eqnarray}
&&\int_{V_{0}}^{V_{1}+1}M_{b}^{a}B^{b}|_{s=-1}\,dv_{-}=L^{-2}\{(v_{-}-\psi
_{+})^{-1}[1,-\mathbf{\hat{y}}-\mathbf{\hat{z}}]^{a}\}|_{V_{0}}^{V_{1}+1}
\nonumber \\
&&-2L^{-2}\{(v_{-}-\psi _{+})^{-1}[0,\sin 2\pi (v_{-}-u_{-})\mathbf{\hat{x}}%
-\cos 2\pi (v_{-}-u_{-})\mathbf{\hat{y}]}^{a}\}|_{V_{0}}^{V_{1}+1}  \nonumber
\\
&&+4\pi L^{-2}\{[0,\{\func{Ci}(2\pi |v_{-}-\psi _{+}|)\cos 2\pi (u_{-}-\psi
_{+})  \nonumber \\
&&-\func{Si}(2\pi |v_{-}-\psi _{+}|)\sin 2\pi (u_{-}-\psi _{+})\}\mathbf{%
\hat{x}}  \nonumber \\
&&+\{\func{Ci}(2\pi |v_{-}-\psi _{+}|)\sin 2\pi (u_{-}-\psi _{+})  \nonumber
\\
&&-\func{Si}(2\pi |v_{-}-\psi _{+}|)\cos 2\pi (u_{-}-\psi _{+})\}\mathbf{%
\hat{y}}]^{a}\}|_{V_{0}}^{V_{1}+1}
\end{eqnarray}
where $\func{Si}(x)=\int_{0}^{x}w^{-1}\sin w\,dw$ and $\func{Ci}%
(x)=-\int_{x}^{\infty }w^{-1}\cos w\,dw$ are again the sine and cosine
integrals, and $\psi _{+}=2\bar{t}/L+2\bar{z}/L$ as before.

In the limit as $[\bar{t},\mathbf{\bar{x}}]\rightarrow [\bar{\tau},\mathbf{%
\bar{X}}]$, therefore,
\begin{eqnarray}
&&\int_{V_{0}}^{V_{1}+1}M_{b}^{a}B^{b}|_{s=-1}\,dv_{-}\rightarrow
L^{-2}\{(2u_{+})^{-1}-D_{+}^{-1}\}[1,-\mathbf{\hat{y}}-\mathbf{\hat{z}}%
]^{a}+L^{-2}u_{+}^{-1}[0,\mathbf{\hat{y}]}^{a}  \nonumber \\
&&+2L^{-2}D_{+}^{-1}[0,\sin 2\pi W_{1}\mathbf{\hat{x}}-\cos 2\pi W_{1}%
\mathbf{\hat{y}]}^{a}  \nonumber \\
&&+4\pi L^{-2}[0,(-\{\func{Ci}(4\pi u_{+})-\func{Ci}(2\pi D_{+})\}\cos 4\pi
u_{+}-\{\func{Si}(4\pi u_{+})-\func{Si}(2\pi D_{+})\}\sin 4\pi u_{+})\mathbf{%
\hat{x}}  \nonumber \\
&&-(\func{Ci}(4\pi u_{+})-\func{Ci}(2\pi D_{+})\}\sin 4\pi u_{+}+\{\func{Si}%
(4\pi u_{+})-\func{Si}(2\pi D_{+})\}\cos 4\pi u_{+})\mathbf{\hat{y}}]^{a}
\label{Bint}
\end{eqnarray}
Collecting together (\ref{A1}), (\ref{A2}), (\ref{A3}) and (\ref{Bint}),
plus the definition (\ref{W1}) of $W_{1}$ in terms of $u_{+}$, gives finally
\begin{equation}
\alpha ^{t}=\alpha ^{z}=32\pi ^{2}\mu L^{-1}F(W_{1})W_{1}(1-\cos 2\pi W_{1})
\label{vert1}
\end{equation}
and
\begin{eqnarray}
&&\left[
\begin{array}{l}
\alpha ^{L} \\
\alpha ^{N}
\end{array}
\right] =-32\pi \mu L^{-1}\times  \nonumber \\
&&\left[
\begin{array}{l}
\{\func{Si}(4\pi u_{+})-\func{Si}(2\pi D_{+})\}\sin 4\pi u_{+}+\{\func{Ci}%
(4\pi u_{+})-\func{Ci}(2\pi D_{+})\}\cos 4\pi u_{+} \\
-\{\func{Si}(4\pi u_{+})-\func{Si}(2\pi D_{+})\}\cos 4\pi u_{+}+\{\func{Ci}%
(4\pi u_{+})-\func{Ci}(2\pi D_{+})\}\sin 4\pi u_{+}
\end{array}
\right]  \nonumber \\
&&+32\pi ^{2}\mu L^{-1}F(W_{1})W_{1}\left[
\begin{array}{l}
2\pi W_{1}\cos 2\pi W_{1}-\sin 2\pi W\allowbreak _{1} \\
\cos 2\pi W_{1}-1+2\pi W_{1}\sin 2\pi W_{1}
\end{array}
\right]
\end{eqnarray}
where
\begin{equation}
F(W)=(2\pi ^{2}W^{2}+1-\cos 2\pi W-2\pi W\sin 2\pi W)^{-1},
\end{equation}
\begin{equation}
4\pi u_{+}=2\pi (1+W_{1})-\tfrac{1}{\pi }(1-\cos 2\pi W_{1})/W_{1}
\end{equation}
and
\begin{equation}
2\pi D_{+}=-\tfrac{1}{\pi }(1-\cos 2\pi W_{1})/W_{1}.
\end{equation}

A similar calculation in the case where $u_{+}<0$ (and so $\bar{s}=-1$)
gives:
\begin{equation}
\alpha ^{t}=-\alpha ^{z}=32\pi ^{2}\mu L^{-1}F(W_{0})W_{0}(1-\cos 2\pi W_{0})
\label{vert2}
\end{equation}
and
\begin{eqnarray}
&&\left[
\begin{array}{l}
\alpha ^{L} \\
\alpha ^{N}
\end{array}
\right] =-32\pi \mu L^{-1}\times  \nonumber \\
&&\left[
\begin{array}{l}
\{\func{Si}(2\pi (1+2u_{+}))-\func{Si}(2\pi D_{+})\}\sin 4\pi u_{+}+\{\func{%
Ci}(2\pi (1+2u_{+}))-\func{Ci}(2\pi D_{+})\}\cos 4\pi u_{+} \\
-\{\func{Si}(2\pi (1+2u_{+}))-\func{Si}(2\pi D_{+})\}\cos 4\pi u_{+}+\{\func{%
Ci}(2\pi (1+2u_{+}))-\func{Ci}(2\pi D_{+})\}\sin 4\pi u_{+}
\end{array}
\right]  \nonumber \\
&&+32\pi ^{2}\mu L^{-1}F(W_{0})W_{0}\left[
\begin{array}{l}
2\pi W_{0}\cos 2\pi W_{0}-\sin 2\pi W\allowbreak _{0} \\
\cos 2\pi W_{0}-1+2\pi W_{0}\sin 2\pi W_{0}
\end{array}
\right]
\end{eqnarray}
where now
\begin{equation}
4\pi u_{+}=2\pi W_{0}-\tfrac{1}{\pi }(1-\cos 2\pi W_{0})/W_{0}
\end{equation}
and
\begin{equation}
2\pi D_{+}=-\tfrac{1}{\pi }(1-\cos 2\pi W_{0})/W_{0}.
\end{equation}

As can be seen from (\ref{vert1}) and (\ref{vert2}), $\alpha ^{t}$ is the
same function $\alpha ^{t}(W)$ of $W_{0}$ for $u_{+}\in (-\frac{1}{2},0)$ as
it is of $W_{1}$ for $u_{+}\in (0,\frac{1}{2})$. Since $%
W_{1}(u_{+})=W_{0}(u_{+}-\frac{1}{2})$ for all $u_{+}\in (0,\frac{1}{2})$
the acceleration component $\alpha ^{t}$ is therefore a periodic function of
$u_{+}$ with period $\frac{1}{2}$. The same is true of the lateral and
normal components $\alpha ^{L}$ and $\alpha ^{N}$, while the vertical
component $\alpha ^{z}$ is anti-periodic.

\section{Deriving the equation of motion from the Battye-Carter equation}

In this appendix I briefly describe the formalism behind the
gauge-independent Battye-Carter equation of motion (\ref{perturb}), and
demonstrate that the equation reduces to the weak-field equations of motion (%
\ref{peqmo0}) and (\ref{peqmo1}) in the standard gauge.

If $(g_{ab},X^{a})$ is a general solution pair to the strong-field
back-reaction problem then the projection tensor $p^{ab}$ corresponding to
the induced 2-metric $\gamma _{AB}=g_{ab}X^{a},_{A}X^{b},_{B}$ is
\begin{equation}
p^{ab}=\gamma ^{AB}X^{a},_{A}X^{b},_{B}
\end{equation}
with $\gamma ^{AB}$ the inverse of $\gamma _{AB}$. The extrinsic curvature
tensor of the world sheet $\mathbf{T}$ is therefore
\begin{equation}
K^{abc}=p_{m}^{a}p^{bn}\nabla _{n}p^{mc}
\end{equation}
where $\nabla _{n}$ is the derivative operator associated with $g_{ab}$.

In terms of the derivatives of the position function $X^{a}$ the curvature
tensor $K^{abc}$ and its trace $K^{c}\equiv g_{ab}K^{abc}$ can be written as
\begin{equation}
K^{abc}=q_{d}^{c}(\gamma ^{AC}\gamma
^{BD}X^{a},_{A}X^{b},_{B}X^{d},_{CD}+p^{am}p^{bn}\Gamma _{mn}^{d})
\end{equation}
and
\begin{equation}
K^{c}=q_{d}^{c}(\gamma ^{CD}X^{d},_{CD}+p^{mn}\Gamma _{mn}^{d})
\end{equation}
with $q^{ab}=g^{ab}-p^{ab}$ the orthogonal complement of $p^{ab}$, and $%
\Gamma _{bc}^{a}$ the Christoffel symbol associated with $g_{ab}$. The
equation of motion of the string, obtained by varying the Lagrangian density
$\mathcal{L}\equiv -\mu \gamma ^{1/2}$, is simply $K^{c}=0$.

If the equation of motion $K^{c}=0$ is perturbed by replacing $g_{ab}$ with $%
g_{ab}^{(0)}+h_{ab}$ then, to linear order in $h_{ab}$,
\begin{equation}
K^{c}-(K^{abc}+K^{a}p^{bc})h_{ab}+q^{cd}p^{ab}(\nabla _{a}h_{bd}-\tfrac{1}{2}%
\nabla _{d}h_{ab})=0  \label{perturb}
\end{equation}
with all the geometrical quantities in this equation, except $h_{ab}$, now
evaluated using $g_{ab}^{(0)}$ in place of $g_{ab}$ \cite{Cart1, Cart2}. In
particular, $K^{c}$ is no longer identically zero but instead of order $%
h_{ab}$. However, the term $K^{a}p^{bc}h_{ab}$ is of second order in $h_{ab}$
and can be ignored.

The Battye-Carter equation (\ref{perturb}) can be specialised to the problem
at hand by first setting $g_{ab}^{(0)}=\eta _{ab}$, so that $\Gamma
_{mn}^{d}=0$. Then to linear order in $h_{ab}$ (\ref{perturb}) reads:
\begin{equation}
q_{d}^{c}(\gamma ^{CD}X^{d},_{CD}-\gamma ^{AC}\gamma
^{BD}h_{ab}X^{a},_{A}X^{b},_{B}X^{d},_{CD})=-q^{cd}\gamma
^{AB}X^{a},_{A}X^{b},_{B}(h_{bd},_{a}-\tfrac{1}{2}h_{ab},_{d})  \label{pert2}
\end{equation}
where $q_{d}^{c}$ and $\gamma ^{AC}$ are of course evaluated using $\eta
_{ab}$ rather than $g_{ab}$.

The next step in the perturbation is to replace $X^{a}$ with $%
X_{(0)}^{a}+\delta X^{a}$, where $\delta X^{a}$ is of linear order in $%
h_{ab} $. The only substantial change is to the term $q_{d}^{c}\gamma
^{CD}X^{d},_{CD}$, which becomes
\begin{equation}
q_{d}^{c}\gamma ^{CD}X_{(0)}^{d},_{CD}+\delta q_{d}^{c}\gamma
^{CD}X_{(0)}^{d},_{CD}+q_{d}^{c}(\gamma ^{CD}\delta X^{d},_{CD}-2\gamma
^{AC}\gamma ^{BD}\eta _{ab}X_{(0)}^{a},_{A}\delta
X^{b},_{B}X_{(0)}^{d},_{CD})
\end{equation}
where $q_{d}^{c}$ and $\gamma ^{AC}$ are now evaluated using $X_{(0)}^{a}$
rather than $X^{a}$, and
\begin{equation}
\delta q_{d}^{c}\equiv -\delta p_{d}^{c}=-2\eta _{db}\gamma
^{AB}q_{a}^{(b}X_{(0)}^{c)},_{A}\delta X^{a},_{B}
\end{equation}
with round brackets on spacetime indices denoting symmetrisation.

Equation (\ref{pert2}) therefore splits into two parts:
\begin{equation}
q_{d}^{c}\gamma ^{CD}X_{(0)}^{d},_{CD}=0  \label{perteqn0}
\end{equation}
and
\begin{eqnarray}
&&q_{d}^{c}\gamma ^{CD}\delta X^{d},_{CD}-2\eta _{db}\gamma
^{AB}q_{a}^{(b}X_{(0)}^{c)},_{A}\delta X^{a},_{B}\gamma
^{CD}X_{(0)}^{d},_{CD}  \nonumber \\
&&-q_{d}^{c}\gamma ^{AC}\gamma ^{BD}(2\eta _{ab}X_{(0)}^{a},_{A}\delta
X^{b},_{B}+h_{ab}X_{(0)}^{a},_{A}X_{(0)}^{b},_{B})X_{(0)}^{d},_{CD}
\nonumber \\
&=&-q^{cd}\gamma ^{AB}X_{(0)}^{a},_{A}X_{(0)}^{b},_{B}(h_{bd},_{a}-\tfrac{1}{%
2}h_{ab},_{d})  \label{perteqn1}
\end{eqnarray}
at zeroth order and first order in $h_{ab}$ respectively.

The final step is to fix the gauge. To keep the analysis as general as
possible, suppose that the full solution pair $(g_{ab},X^{a})$ satisfies a
gauge condition of the form
\begin{equation}
\gamma _{AB}=\gamma ^{1/2}\kappa _{AB}
\end{equation}
where $\kappa _{AB}$ is a constant, symmetric $2\times 2$ matrix independent
of $\mu $, with determinant $-1$. The standard gauge, for example, has $%
\kappa _{AB}=\eta _{AB}$. Replacing $g_{ab}$ with $\eta _{ab}+h_{ab}$ and $%
X^{a}$ with $X_{(0)}^{a}+\delta X^{a}$ then gives a zeroth-order gauge
constraint
\begin{equation}
\gamma ^{-1/2}\gamma _{AB}=\kappa _{AB}  \label{const0}
\end{equation}
and a first-order gauge constraint
\begin{eqnarray}
0 &=&\Delta \kappa _{AB}\equiv \gamma ^{-1/2}[2\eta
_{ab}X_{(0)}^{a},_{(A}\delta
X^{b},_{B)}+h_{ab}X_{(0)}^{a},_{A}X_{(0)}^{b},_{B}  \nonumber \\
&&-\gamma ^{CD}(\eta _{ab}X_{(0)}^{a},_{C}\delta X^{b},_{D}+\tfrac{1}{2}%
h_{ab}X_{(0)}^{a},_{C}X_{(0)}^{b},_{D})\gamma _{AB}]  \label{const1}
\end{eqnarray}
where, as before, all geometric quantities are evaluated using $\eta _{ab}$
and $X_{(0)}^{a}$ in place of $g_{ab}$ and $X^{a}$.

If the zeroth-order constraint (\ref{const0}) is substituted into the
zeroth-order equation of motion (\ref{perteqn0}) then, since $%
q_{d}^{c}=\delta _{d}^{c}-p_{d}^{c}$ and
\begin{equation}
p_{d}^{c}\gamma ^{CD}X_{(0)}^{d},_{CD}=-\gamma ^{-1/2}(\gamma ^{1/2}\gamma
^{CD}),_{D}X_{(0)}^{c},_{C}  \label{identity}
\end{equation}
it follows that
\begin{equation}
0=q_{d}^{c}\gamma ^{CD}X_{(0)}^{d},_{CD}=\gamma ^{-1/2}(\gamma ^{1/2}\gamma
^{CD}X_{(0)}^{c},_{C}),_{D}=\gamma ^{CD}X_{(0)}^{c},_{CD}
\end{equation}
as $\gamma ^{1/2}\gamma ^{CD}=\kappa ^{CD}$, the inverse of $\kappa _{CD}$.
The zeroth-order equation of motion is therefore simply $\gamma
^{CD}X_{(0)}^{d},_{CD}=0$, which in the standard gauge reduces to (\ref
{peqmo0}) as required. It might seem odd that the equation $q_{d}^{c}\gamma
^{CD}X_{(0)}^{d},_{CD}=0$, which contains only two algebraically-independent
components, should be equivalent to an equation $\gamma
^{CD}X_{(0)}^{d},_{CD}=0$ with four independent components. However, as is
evident from (\ref{identity}), the equation $\gamma ^{CD}X_{(0)}^{d},_{CD}=0$
contains information about the gauge choice as well as the equation of
motion.

The first-order equation of motion (\ref{perteqn1}) now reduces to
\begin{eqnarray}
&&q_{d}^{c}\gamma ^{CD}\delta X^{d},_{CD}-q_{d}^{c}\gamma ^{AC}\gamma
^{BD}(2\eta _{ab}X_{(0)}^{a},_{A}\delta
X^{b},_{B}+h_{ab}X_{(0)}^{a},_{A}X_{(0)}^{b},_{B})X_{(0)}^{d},_{CD}
\nonumber \\
&=&-q^{cd}\gamma ^{AB}X_{(0)}^{a},_{A}X_{(0)}^{b},_{B}(h_{bd},_{a}-\tfrac{1}{%
2}h_{ab},_{d})
\end{eqnarray}
where
\begin{eqnarray}
&&\gamma ^{AC}\gamma ^{BD}(2\eta _{ab}X_{(0)}^{a},_{A}\delta
X^{b},_{B}+h_{ab}X_{(0)}^{a},_{A}X_{(0)}^{b},_{B})X_{(0)}^{d},_{CD}
\nonumber \\
&=&\gamma ^{1/2}\gamma ^{AC}\gamma ^{BD}\Delta \kappa
_{AB}X_{(0)}^{d},_{CD}+\gamma ^{EF}(\eta _{ab}X_{(0)}^{a},_{E}\delta
X^{b},_{F}+\tfrac{1}{2}h_{ab}X_{(0)}^{a},_{E}X_{(0)}^{b},_{F})\gamma
^{CD}X_{(0)}^{d},_{CD}  \nonumber \\
&=&0
\end{eqnarray}
as $\Delta \kappa _{AB}=0$ and $\gamma ^{CD}X_{(0)}^{d},_{CD}=0$. So the
first-order equation of motion becomes
\begin{equation}
q_{d}^{c}\gamma ^{CD}\delta X^{d},_{CD}=-q^{cd}\gamma
^{AB}X_{(0)}^{a},_{A}X_{(0)}^{b},_{B}(h_{bd},_{a}-\tfrac{1}{2}h_{ab},_{d}).
\label{perteqn2}
\end{equation}

Finally, after considerable algebraic rearrangement it can be seen that
\begin{eqnarray}
&&p_{d}^{c}\gamma ^{CD}\delta X^{d},_{CD}+p^{cd}\gamma
^{AB}X_{(0)}^{a},_{A}X_{(0)}^{b},_{B}(h_{bd},_{a}-\tfrac{1}{2}h_{ab},_{d})
\nonumber \\
&=&\gamma ^{1/2}\gamma ^{AB}X_{(0)}^{c},_{B}\gamma ^{CD}(\Delta \kappa
_{AC}),_{D}  \nonumber \\
&&-\gamma ^{AB}X_{(0)}^{c},_{B}(h_{ab}X_{(0)}^{c},_{A}+\eta _{ab}\delta
X^{a},_{A})\gamma ^{CD}X_{(0)}^{d},_{CD}  \nonumber \\
&&-\gamma ^{1/2}\gamma ^{AB}X_{(0)}^{c},_{B}\gamma ^{CD}\gamma ^{EF}\eta
_{ab}X_{(0)}^{a},_{E}(X_{(0)}^{b},_{AD}\Delta \kappa
_{CF}-X_{(0)}^{b},_{FD}\Delta \kappa _{CA})  \nonumber \\
&&+\gamma ^{1/2}\gamma ^{AB}X_{(0)}^{c},_{B}\gamma ^{EF}X_{(0)}^{a},_{E}(%
\tfrac{1}{2}h_{ab}X_{(0)}^{c},_{F}+\eta _{ab}\delta X^{a},_{F})\gamma
^{CD}(\gamma ^{-1/2}\gamma _{AC}),_{D}  \nonumber \\
&=&0  \label{tange}
\end{eqnarray}
(again because $\gamma ^{-1/2}\gamma _{AB}$ is constant, $\Delta \kappa
_{AB}=0$ and $\gamma ^{CD}X_{(0)}^{d},_{CD}=0$), and so, given that $%
q^{cd}=\eta ^{cd}-p^{cd}$, the first-order equation of motion (\ref{perteqn2}%
) takes the form
\begin{equation}
\gamma ^{CD}\delta X^{c},_{CD}=-\eta ^{cd}\gamma
^{AB}X_{(0)}^{a},_{A}X_{(0)}^{b},_{B}(h_{bd},_{a}-\tfrac{1}{2}h_{ab},_{d})
\label{perteqn3}
\end{equation}
which in the standard gauge is just (\ref{peqmo1}). As with the zeroth-order
equation of motion, the fact that (\ref{perteqn3}) has four independent
components does not mean that the system of equations is over-determined, as
the tangential components (\ref{tange}) only contain information about the
choice of gauge.

Conversely, if the zeroth-order gauge condition (\ref{const0}) is imposed
then
\begin{eqnarray}
\kappa ^{BC}(\gamma ^{1/2}\Delta \kappa _{AB}),_{C} &=&\gamma ^{1/2}\eta
_{ce}X_{(0)}^{e},_{A}[\gamma ^{CD}\delta X^{c},_{CD}+\eta ^{cd}\gamma
^{AB}X_{(0)}^{a},_{A}X_{(0)}^{b},_{B}(h_{bd},_{a}-\tfrac{1}{2}h_{ab},_{d})]
\nonumber \\
&&+\gamma ^{1/2}(\eta _{cd}\delta X^{c},_{A}+h_{cd}X_{(0)}^{c},_{A})\gamma
^{CD}X_{(0)}^{d},_{CD}
\end{eqnarray}
and so if $X_{(0)}^{a}$ and $\delta X^{a}$ satisfy the zeroth- and
first-order equations of motion $\gamma ^{CD}X_{(0)}^{d},_{CD}=0$ and (\ref
{perteqn3}), the evolution of the first-order constraint matrix $\Delta
\kappa _{AB}$ is governed by the simple system of equations $\kappa
^{BC}(\gamma ^{1/2}\Delta \kappa _{AB}),_{C}=0$.

To analyse this system, first rotate the gauge coordinates $\zeta ^{A}$ so
that the symmetric matrix $\kappa ^{BC}$ is diagonal, by setting $\zeta
^{A^{\prime }}=\Lambda _{A}^{A^{\prime }}\zeta ^{A}$ where the orthogonal
matrix $\Lambda _{A}^{A^{\prime }}$ is defined by
\begin{equation}
\Lambda _{B}^{B^{\prime }}\Lambda _{C}^{C^{\prime }}\kappa ^{BC}\equiv
\kappa ^{B^{\prime }C^{\prime }}=\text{diag}(\lambda _{1},\lambda _{2}).
\end{equation}
Then the function $U_{A^{\prime }B^{\prime }}=(\Lambda _{B}^{B^{\prime
}}\Lambda _{A}^{A^{\prime }})^{-1}\gamma ^{1/2}\Delta \kappa _{AB}$
satisfies the equation $\kappa ^{B^{\prime }C^{\prime }}U_{A^{\prime
}B^{\prime }},_{C^{\prime }}=0$. Furthermore, the matrix $\Delta \kappa
_{AB} $ is by definition (\ref{const1}) trace-free, and so
\begin{equation}
\kappa ^{A^{\prime }B^{\prime }}U_{A^{\prime }B^{\prime }}=\gamma
^{1/2}\kappa ^{AB}\Delta \kappa _{AB}=0.
\end{equation}
On setting $u=\frac{1}{2}(\lambda _{1}U_{1^{\prime }1^{\prime }}-\lambda
_{2}U_{2^{\prime }2^{\prime }})$ and $v=U_{1^{\prime }2^{\prime }}$ the
equation $\kappa ^{B^{\prime }C^{\prime }}U_{A^{\prime }B^{\prime
}},_{C^{\prime }}=0$ reduces to the linear system
\begin{equation}
A\mathbf{u},_{1^{\prime }}+B\mathbf{u},_{2^{\prime }}=0
\end{equation}
where
\begin{equation}
A=\left[
\begin{array}{ll}
1 & 0 \\
0 & \lambda _{1}
\end{array}
\right] \text{,\qquad }B=\left[
\begin{array}{ll}
0 & \lambda _{2} \\
-1 & 0
\end{array}
\right] \qquad \text{and\qquad }\mathbf{u}=(u,v)^{T}.
\end{equation}

The classification and characteristics of this system of equations are
determined by the roots $\xi =\pm \sqrt{-\lambda _{1}/\lambda _{2}}$ of the
equation $\det (A-\xi B)=0$. Since the induced metric $\gamma ^{AB}=\gamma
^{-1/2}\kappa ^{AB}$ has signature $(+,-)$, the roots $\xi $ are real and
distinct, and the system of equations is hyperbolic. Moreover, the
characteristics satisfy the equation $d\zeta ^{1^{\prime }}/d\zeta
^{2^{\prime }}=\xi $, and so have tangent directions $t^{A^{\prime }}=[\xi
,1]$. Hence,
\begin{equation}
\kappa _{AB}t^{A}t^{B}=\kappa _{A^{\prime }B^{\prime }}t^{A^{\prime
}}t^{B^{\prime }}=\xi ^{2}\lambda _{1}^{-1}+\lambda _{2}^{-1}=0
\end{equation}
and the characteristics are null curves on the world sheet. The equation $%
\kappa ^{BC}(\gamma ^{1/2}\Delta \kappa _{AB}),_{C}=0$ therefore has the
unique solution $\gamma ^{1/2}\Delta \kappa _{AB}=0$ (and satisfies the
first-order gauge constraint (\ref{const1})) at all points on the world
sheet, provided that $\Delta \kappa _{AB}=0$ on the non-null initial curve $%
\tau =0$.

It is always possible to ensure that $\Delta \kappa _{AB}=0$ on the initial
curve by choosing the initial data $\delta X^{a},_{A}$ so that (\ref{const1}%
) is satisfied at all points on the curve. The details of this procedure as
it applies to the ACO loop are explained at the beginning of Section \ref
{evolsec}. In the present case, the above discussion is modified slightly by
the presence of the kinks, because $\gamma _{AB}$ is undefined (that is,
discontinuous) at the kink points, and so strictly speaking are $\kappa
_{AB} $ and $\Delta \kappa _{AB}$ (although continuous extensions of these
functions exist trivially). Nonetheless, as is shown explicitly in Section
\ref{evolsec}, the first-order constraint $\Delta \kappa _{AB}=0$ is still
satisfied at all ordinary points on the world sheet.

\newpage

LIST\ OF\ FIGURE CAPTIONS:\bigskip

Figure 1: $y$-$z$ projection of the Allen-Casper-Ottewill loop with minimum
radiative efficiency.\bigskip

Figure 2: Backwards light cones for a field point ($A$) located at the kink
point $u_{+}=u_{-}=0$, and for a general field point ($B$) on the world
sheet.\bigskip

Figure 3: The time component $\alpha ^{t}$ of the acceleration vector, in
units of $\mu /L$, as a function of $u_{+}$. The $z$ component $\alpha ^{z}$
is identical for $u_{+}>0$, but equals $|\alpha ^{t}|$ for $u_{+}<0$.\bigskip

Figure 4: The lateral component $\alpha ^{L}$ of the acceleration vector, in
units of $\mu /L$, as a function of $u_{+}$.\bigskip

Figure 5: The normal component $\alpha ^{N}$ of the acceleration vector, in
units of $\mu /L$, as a function of $u_{+}$.

\newpage

\begin{figure}
\epsfig{file=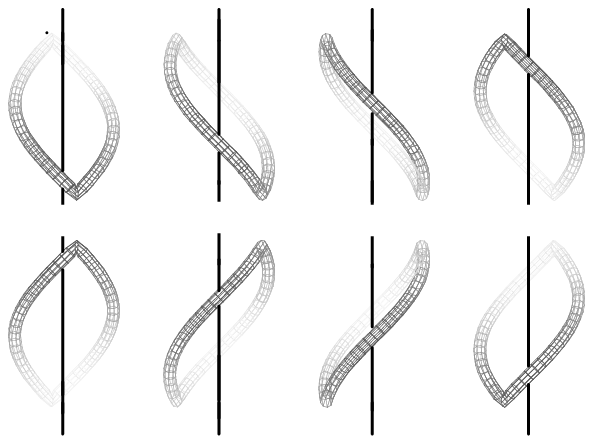, height=1.8in}
\end{figure}
FIGURE 1


\newpage

\begin{figure}
\epsfig{file=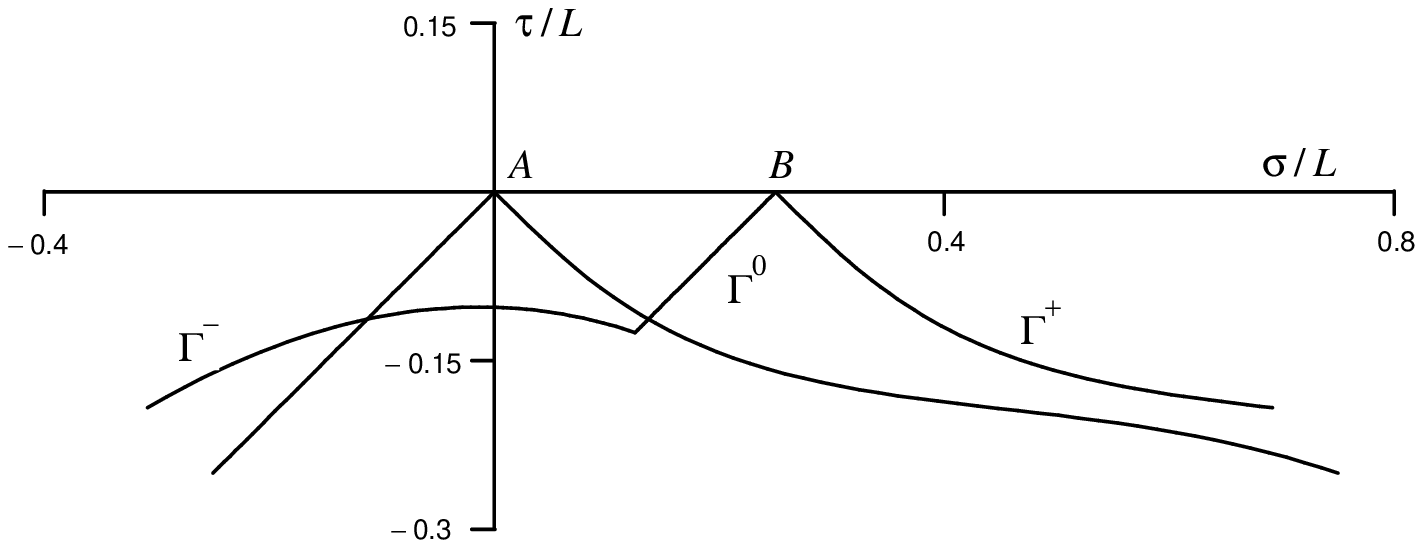, height=1.8in}
\end{figure}
FIGURE 2


\newpage

\begin{figure}
\epsfig{file=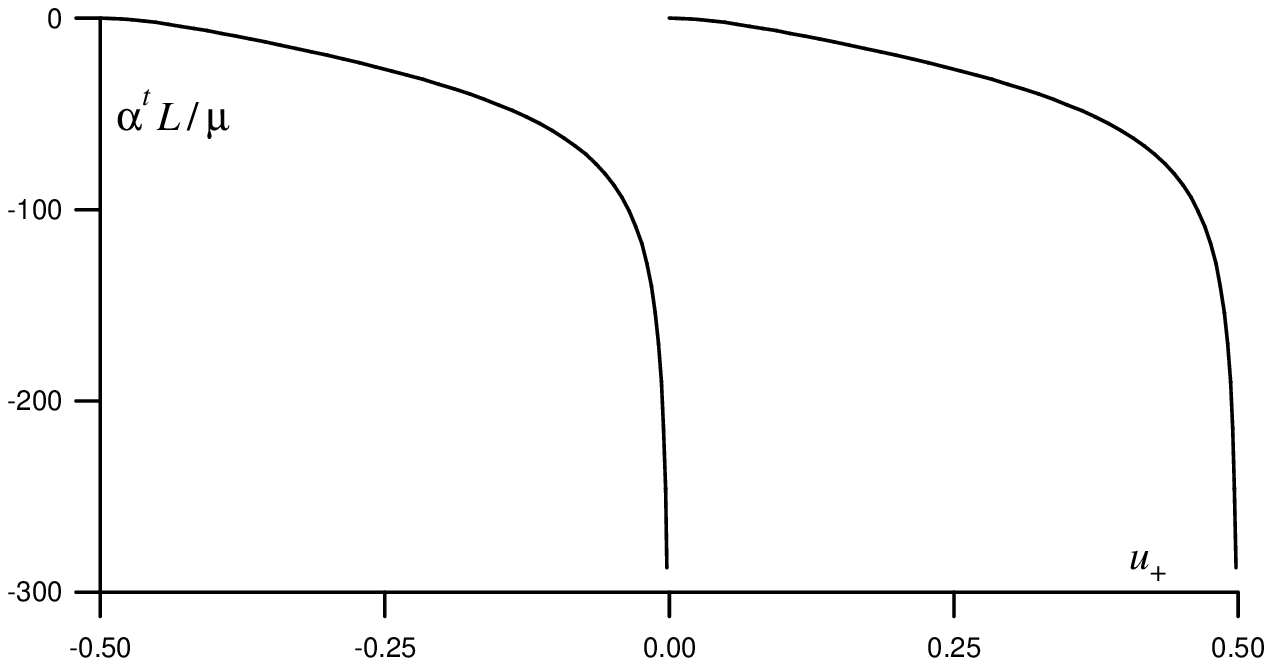, height=1.8in}
\end{figure}
FIGURE 3


\newpage

\begin{figure}
\epsfig{file=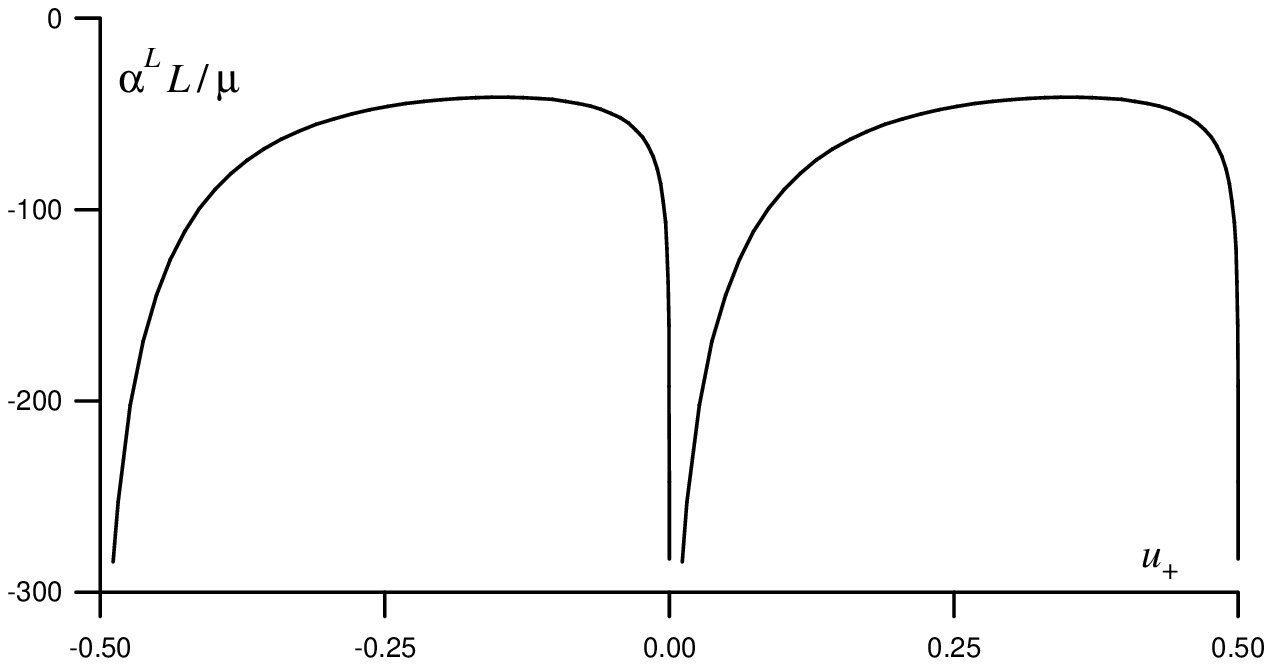, height=1.8in}
\end{figure}
FIGURE 4


\newpage

\begin{figure}
\epsfig{file=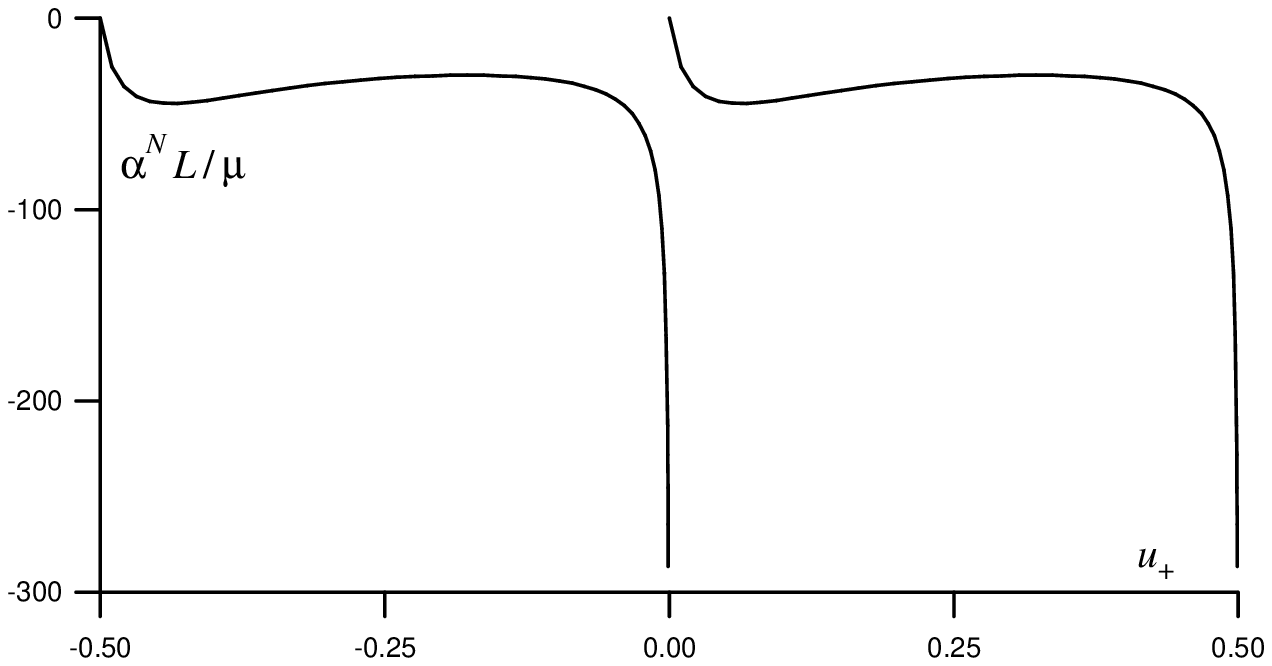, height=1.8in}
\end{figure}
FIGURE 5


\end{document}